\DocumentMetadata{%
	pdfstandard=A-3b,
	pdfversion=1.7,
	lang=en-US,
}
\documentclass[barcolor=Goldenrod3,mathalfa=cal=euler,varvw,upint,subscriptcorrection,balance,nolists,nofoot]{asmejour}

\usepackage{soul}

\newtheorem{theorem}{Theorem}

\newcommand*{\eu}{\mathord{e}}
\DeclareMathSymbol{\mh}{\mathord}{operators}{`\-} 

\DeclareMathOperator{\re}{Re}
\DeclarePairedDelimiterXPP\myRe[1]{\re\mkern2mu}\{\}{}{#1}

\DeclareMathOperator{\im}{Im}
\DeclarePairedDelimiterXPP\myIm[1]{\im\mkern2mu}\{\}{}{#1}

\DeclarePairedDelimiter\abs{\lvert}{\rvert} 

\hypersetup{%
	pdfauthor={Kyle I. McKee and John H. Lienhard}, 
    pdftitle={Symmetry criteria for the equality of interior and exterior shape factors},  
	pdfkeywords={heat conduction, shape factor, interior, exterior, symmetry, conformal mapping},
	pdfsubject = {Theorems on shape factors inside and outside planar objects},	
 }

\JourName{Heat and Mass Transfer}

\begin{document}

\title{Symmetry criteria for the equality of interior and exterior shape factors}

\SetAuthorBlock{Kyle I. McKee}{Department of Mathematics,\\ Massachusetts Institute of Technology,\\
Cambridge, MA 02139 USA\\[4pt]
Gulliver Laboratory,\\
ESPCI Paris, PSL University,\\
75231 Paris cedex 05, France\\
email: kimckee@mit.edu}
\SetAuthorBlock{John H. Lienhard\CorrespondingAuthor}{Fellow ASME\\
Rohsenow Kendall Heat Transfer Laboratory,\\
Department of Mechanical Engineering,\\
Massachusetts Institute of Technology,\\
Cambridge, MA 02139 USA\\
email: lienhard@mit.edu
}


\begin{abstract}
Lienhard~\cite{lienhard2019exterior} reported that the shape factor of the interior of a simply-connected region ($\Omega$) is equal to that of its exterior ($\mathbb{R}^2\backslash\Omega$) under the same boundary conditions. In that study, numerical examples supported the claim in particular cases; for example, it was shown that for certain boundary conditions on circles and squares, the conjecture holds. In the present paper, we show that the conjecture is not generally true, unless some additional condition is met.  We proceed by elucidating why the conjecture does in fact hold in all of the examples analysed by Lienhard.  We thus deduce a simple criterion which, when satisfied, ensures the equality of interior and exterior shape factors in general. Our criterion notably relies on a beautiful and little-known symmetry method due to Hersch~\cite{hersch1982harmonic} which we introduce in a tutorial manner. 
\end{abstract}

\maketitle


\section{Introduction}\label{sec:intro}
We consider steady heat conduction inside or outside a planar object. The object has different sections of its boundary at one of two temperatures, with the isothermal sections separated from one another by adiabatic sections. Heat $Q$ [W/m] then flows from the high temperature portion of the boundary to the low temperature portion of the boundary at a rate
\begin{equation}
Q = S k \Delta T
\end{equation}
where $S$ is the shape factor, $\Delta T$ is the temperature difference, and $k$ [W/m-K] is the object's thermal conductivity. The shape factor is dimensionless and scale invariant, and determined exclusively by the particular object's geometry and  boundary conditions~\cite{lienhard2019exterior,yovanovich1998,ahtt5p}. 

The shape factor can be found by solving the Laplace equation in the domain of interest, and then integrating the heat flux exiting the high temperature portion of the boundary. The shape factor conveniently summarizes the result of such analyses. In most cases, the shape factor inside a two-dimensional object is easier to compute (analytically or numerically) than the shape factor outside a two-dimensional object, even for an apparently simple object like a square (see, e.g., \cite[\S7.2]{lienhard2023potential}). Thus, a relationship between the interior and exterior shape factors allows a potentially difficult exterior calculation to be replaced by a more tractable interior calculation or an already known result.

The present paper examines the shape factor in two dimensions using techniques from complex analysis and conformal mapping.  To proceed mathematically, we precisely define the ``object'' boundary as a \textit{Jordan curve}: a closed non-self-intersecting curve that divides the plane into an interior region and an unbounded exterior region. This boundary is partitioned into a finite number of sections, each  prescribed a boundary condition that is either adiabatic (a zero Neumann condition) or isothermal (a constant Dirichlet condition) at one of two temperatures.  Sections with different temperatures must be separated by an adiabatic section to ensure a finite rate of heat transfer. However, any number of sections may be at each temperature.

To mathematicians, the shape factor is the reciprocal of the \emph{extremal distance}---a conformal invariant of interest in function theory that is equal to the reciprocal of the Dirichlet energy\footnote{For a harmonic function $\phi$, defined over a domain $D\subseteq \mathbb{R}^2$, the Dirichlet energy is defined by the following integral over the domain area: $\int_D\left(\nabla \phi\boldsymbol{\cdot}\nabla \phi \right) dA$.}~\cite[pg.~65]{ahlfors2010conformal}. Useful properties of the extremal distance are outlined by Ahlfors~\cite[pg.~48]{ahlfors2010conformal}. When the boundary curve is split into precisely four segments, the extremal distance is equal to the conformal modulus, which was studied by Hersch \cite{hersch1982harmonic, hersch1987mapping} in geometries possessing certain symmetries. By exploiting the Schwarz reflection principle, Hersch derived analytical results for a variety of geometries.

More recently, Lienhard~\cite{lienhard2019exterior} made the striking claim that interior and exterior shape factors are always equal. To reach this conclusion, he first used the Riemann mapping theorem to argue that the interior and exterior of a Jordan curve may always be related by a conformal map. Indeed, such a map is guaranteed to exist~\cite{gamelin2003complex}. 

Lienhard proceeded by further assuming that each boundary section possessing a given boundary condition
is invariant under the interior-to-exterior mapping (up to an overall reflection); we label this assumption~A. When true, assumption A does indeed imply that the interior and exterior shape factor problems are conformally equivalent (see \S \ref{sec:conf-eq}), and thus they possess the same shape factor, $S_i=S_e$ \cite{ahlfors2010conformal}.
Based on assumption~A, Lienhard~\cite{lienhard2019exterior} conjectured that $S_i=S_e$ must always hold. He subsequently analysed the shape factors for a disc and a square, under various boundary conditions, and found in each case that $S_i = S_e$, thus gaining support for the conjecture. However, we now demonstrate that $S_i = S_e$ need not hold in more general cases than those considered by Lienhard.

\begin{figure}
    \centering
    \includegraphics*[width=0.75\columnwidth]{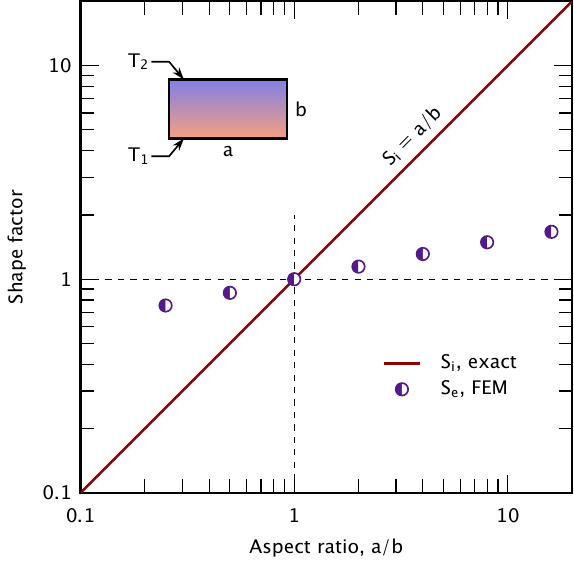}
    \caption{Interior and exterior shape factor for a rectangle of aspect ratio $a/b$. The interior shape factor is exactly $a/b$. The exterior values are from FEM calculations (see Appendix~\ref{sec:FEM-conv}). The interior and exterior shape factors are equal only for the case of a square ($a=b$) but unequal for rectangles with $a \neq b$.}
    \label{fig:rectangle-S}
\end{figure}

Consider the following counter-example: a rectangle with isothermal sides of width $a$ and adiabatic sides with height $b$. For the case of a square, when $a = b$, it is clear that $S_i = S_e = 1$ as was demonstrated previously~\cite{lienhard2019exterior}. However for rectangles, where $a\neq b$, the two shape factors diverge rapidly, as seen in Fig.~\ref{fig:rectangle-S}.

The purpose of the present paper is twofold. First, we shall elucidate why Lienhard's conjecture is not universally true. Second, we provide an additional criterion that does ensure the equality of interior and exterior shape factors. The new criterion relies on a little-known method due to Hersch \cite{hersch1987mapping}, and we thus take this opportunity to explain that method in an expository manner.


\section{Conformally Equivalent Shape Factor Problems}\label{sec:conf-eq}
The failure of assumption~A ultimately spoils the conformal equivalence between the interior and exterior problems, as we now describe.

Consider two Jordan curves $\partial C$ and $\partial D$. 
Then let $C$ be the domain defined as either the interior or exterior of $\partial C$. Let $\partial D$ be the domain defined as either the interior or exterior of $\partial D$. Suppose that each boundary curve is partitioned into a finite number of sections with each section assigned either a zero Neumann condition or a constant Dirichlet condition at one of two temperatures. Again, sections at different temperatures must be separated by an adiabatic segment to ensure a finite rate of heat transfer.

We say that the two shape factor problems, in the domains $C$ and $D$, are conformally equivalent if the following criteria hold: 1) the two problem domains are related by a conformal map; and 2) under that mapping,
the boundary condition at every image point matches that at the corresponding pre-image point. If two shape factor problems are conformally equivalent, then they possess the same shape factor~\cite{ahlfors2010conformal}. The case of interest in the present paper is that for which $C$ is the interior and $D$ the exterior of the \textit{same} Jordan curve.

The Riemann Mapping Theorem guarantees the existence of a conformal map $f(z)$ between the interior and exterior of a Jordan curve, thus guaranteeing the satisfaction of criterion~1. However, the mapping theorem only offers three degrees of freedom: one can always construct a map that sends \textit{three} points on the original domain boundary to three chosen points on the target domain's boundary~\cite[pg.~290]{gamelin2003complex}. 
Note that the choice of the three image points is not completely arbitrary: their order along the boundary must be preserved under the mapping.  Once the image locations of three pre-image points are chosen, the map becomes fully specified so that the location of any other point cannot be chosen at will. 

Shape factors, however, are determined by a minimum of \textit{four} points, one at the end of each isothermal section. As a result, Riemann's theorem does not guarantee that the boundary conditions are preserved when mapping from the interior to the exterior domain. It follows that the interior and exterior boundary value problems are not necessarily conformally equivalent, meaning that the interior and exterior shape factors are not necessarily equal. An additional constraint or symmetry must be present to ensure that the map leaves the boundary conditions invariant, as will be discussed in \S \ref{sec:syms}.
By chance, the specific cases examined by Lienhard each tacitly embodied such symmetries, which led to the incorrect deduction that $S_i=S_e$ must always hold.


\section{A Non-trivial Class of Geometries with Equal Interior and Exterior Shape Factors}\label{sec:syms}
In this section, we apply a powerful method for analysing Laplace problems that was described by Hersch 
in another context~\cite{hersch1987mapping}. The method builds upon the Schwarz reflection principle. By application of this method, we deduce a simple criterion that ensures $S_i=S_e$. 

After assuming that the domain boundary $\partial B$ (Jordan curve) possesses a certain reflectional symmetry, we proceed to construct a conformal map that translates the interior shape factor problem, defined on $\mathrm{int}\left(\partial B\right)$, into a shape factor problem interior to the unit disk. We similarly translate the exterior shape factor problem into a problem interior to the unit disk. Under further symmetry conditions on the boundary condition placements, we demonstrate that both shape factor problems (interior and exterior to $\partial B$) map to the same problem on the interior of the unit circle, and thus possess identical shape factors.

\subsection{Schwarz Reflection Principle: Hersch Sector Reflections and the Interior Problem}\label{intprob}
The Schwarz reflection principle may be stated as follows (\cite{gamelin2003complex}, pg.~283). 
\begin{theorem}(\textbf{Schwarz Reflection})\label{thm:schref}
Consider a domain $W$ symmetric with respect to the real axis, where we define $W^+=W\cap \{\myIm{z}>0\}$ and $W^-=W\cap \{\myIm{z}<0\}$ and $z\in\mathbb{C}$. 
Let $f(z)$ be an analytic function defined on $W^-$ satisfying $\myIm{f(z)}\rightarrow 0$ as $z\rightarrow W\cap\mathbb{R}$. Then $f(z)$ extends to be analytic on $W$ such that
\begin{equation}\label{eq:sreqconj}
f(\overline{z})=\overline{f(z)}.
\end{equation}
\end{theorem}

\begin{figure}[!tb]
    \centering
    \includegraphics*[width=\columnwidth]{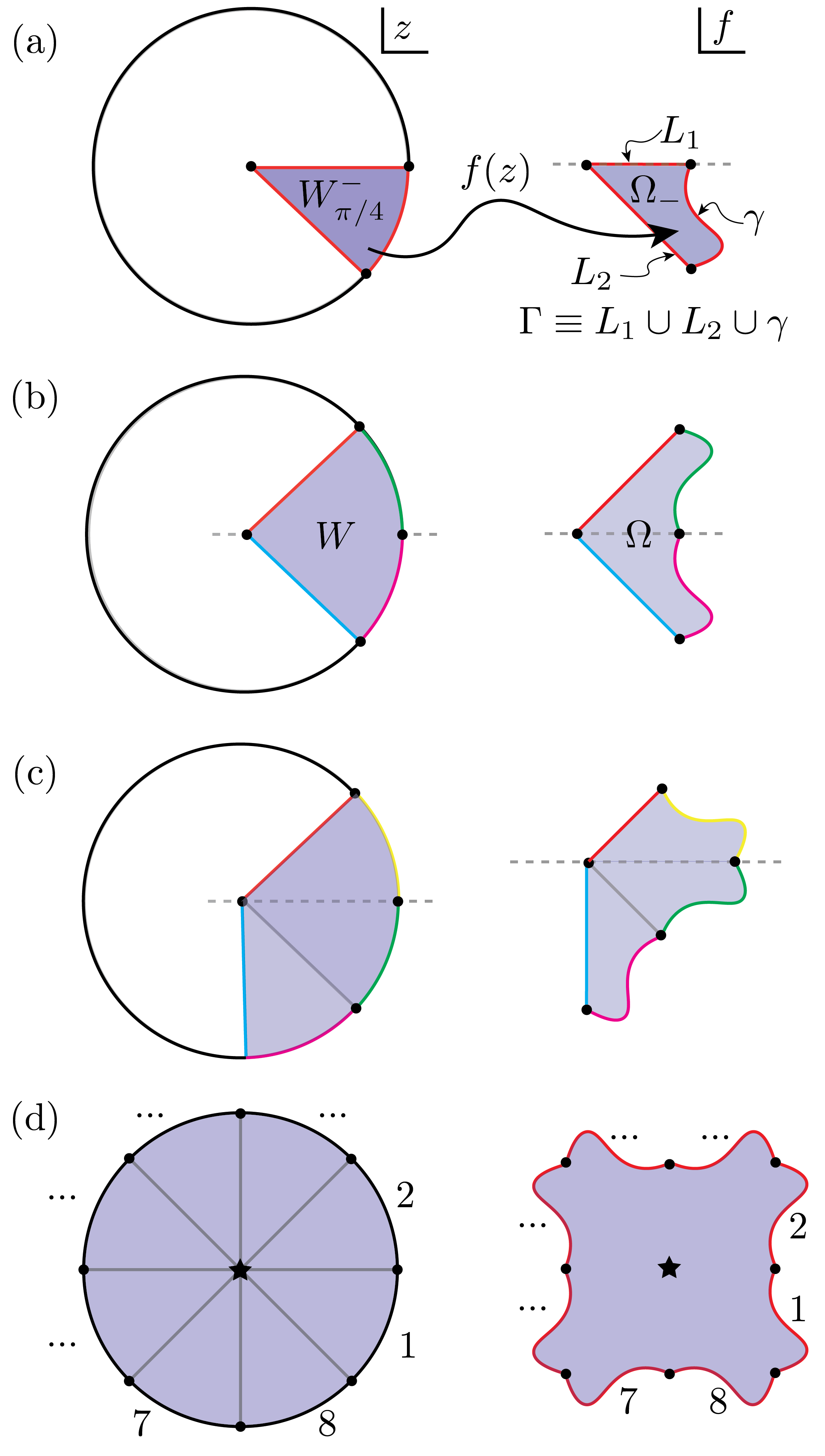}
    \caption{Demonstration of the sector reflection principle of Hersch \cite{hersch1987mapping}. The Riemann mapping theorem guarantees the existence of an analytic mapping $f(z)$ between $W^-_{\pi/4}$ and $\Omega_-$ as indicated in panel (a). The Schwarz reflection principle guarantees $f(z)$ may be analytically extended to $W$ with the boundary correspondence as indicated in panel (b). Rotating clockwise by $\pi/4$  and reflecting again leads to panel (c).
    By performing rotations and reflections, the domain of analyticity of $f(z)$ may be extended to the entire unit circle. The image and boundary correspondence of the mapping from the unit circle is given in panel (d).}
    \label{fig:herschrefl}
\end{figure}

We now illustrate how the Schwarz reflection principle applies to conformal maps of sectors of the complex plane.
Consider a circular sector occupying $\pi/4$ radians of the interior of the unit circle, which we denote $W_{\pi/4}^{-}$, as depicted in Fig.~\ref{fig:herschrefl}(a). The Riemann mapping theorem guarantees that this sector can be transformed into any other simply-connected region by a conformal map. In particular, the mapping theorem ensures that there exists a mapping function $f(z)$ to the region $\Omega_-$ depicted on the right side of Fig.~\ref{fig:herschrefl}(a), whose boundary we denote as $\Gamma$. 

Our analysis applies generally to situations where the boundary curve $\Gamma$ has the following properties. Consider a wedge sector of the complex plane spanning some angle, with its boundaries defined by two straight lines emanating from the origin. Now choose one point on each sector boundary and call the straight segments from the origin to each of these two points \st{as} $L_1$ and $L_2$. Now connect the open endpoints of $L_1$ and $L_2$ to one another with a non-self-intersecting curve $\gamma$ which only intersects the sector boundaries at its endpoints.  We call the outer curve, $\gamma$, a \emph{primitive edge}. Then, we define the \emph{primitive curve}, $\Gamma \equiv L_1 \cup L_2 \cup \gamma$.
We call the shape bounded by $\Gamma$, $\mathrm{int}\left(\Gamma\right)$,
the \emph{primitive shape}. Primitive shapes are shaded in purple in all figures herein.

In addition, we can choose the image location of three pre-image points under the mapping $f(z)$, all of which are labelled as black circles in Fig.~\ref{fig:herschrefl}(b). Then, $f(z)$ maps the domain $W_{\pi/4}^-$ onto its image $\Omega_-$ and maps the three pre-image points to the three image points. 

Now consider the region $W = W_{\pi/4}^+ \cup W_{\pi/4}^-$, in the pre-image domain, which is symmetric about the real axis and illustrated on the left of Fig.~\ref{fig:herschrefl}(b). It is clear that $W$ meets the criterion of the Schwarz reflection principle, and so it is guaranteed that the domain of analyticity of $f(z)$ may be extended from $W^-_{\pi/4}$ to $W$. Moreover, it is guaranteed that the analytic extension of $f(z)$ defined over $W$ satisfies Eq.~\eqref{eq:sreqconj}. It is thus clear that the image of $W$, under the analytic extension of $f(z)$, is $\Omega=\Omega_-\cup\Omega_+$, as drawn on the right of Fig.~\ref{fig:herschrefl}(b). 

To extend the mapping domain further, we first rotate the pre-image $W$ from Fig.~\ref{fig:herschrefl}(b) about the origin by $\pi/4$ radians clockwise: specifically, the new map is written in terms of the old as $\exp(-\mathrm{i}\pi/4)\mkern1.5mu f\big(\exp(\mathrm{i}\pi/4)z\big)$. We next apply the Schwarz reflection principle about the real axis---again taking $W_-$ to be a sector of $\pi/4$ radians in the unit circle (below the real axis) and $\Omega_-$ to be one primitive sector below the real axis---from which we obtain the extended map illustrated in Fig.~\ref{fig:herschrefl}(c). At this point, we have constructed a map that takes a wedge spanning $3\pi/4$ radians in the unit circle to an object containing three primitive shapes (see Fig.~\ref{fig:herschrefl}(c)).

Repeating the procedure of a rotation followed by a sector reflection, one may achieve a mapping of the full unit circle onto the shape in the right panel of Fig.~\ref{fig:herschrefl}(d). Generally, if a body is composed of $N$ primitive shapes, one must perform $N-1$ sector reflections of the primitive shape to construct a map from the unit circle. For example, a total of seven reflections are used to generate the map in Fig.~\ref{fig:herschrefl}(d). Note the geometric requirement that $N$ be even.\looseness=-1

In Fig.~\ref{fig:herschrefl}(d), the equal length boundary segments labelled 1 through 8 on the unit circle boundary are mapped by $f(z)$ to the corresponding labelled segments in the right panel. Despite the fact that the Riemann mapping theorem generally only guarantees three degrees of freedom in a conformal map, 
symmetry has allowed us to fix the location of \textit{nine} points: eight boundary points and the origin in Fig.~\ref{fig:herschrefl}(d).

More generally, if a Jordan curve $\partial B$ may be constructed from $N$ sector reflections of a primitive edge, then there exists a conformal map from the interior of the unit circle to $B\equiv\mathrm{int}(\partial B)$ which maps each of $N$ equal length arcs on the unit circle to a primitive edge in $\partial B$.

It follows that if boundary conditions are specified (and unchanging) along each of the $N$ primitive edges in $\partial B$, then the shape factor problem in $B$ is conformally equivalent to the problem interior to the unit circle having the same boundary conditions specified on $N$ arcs of equal length. The conformal transformation between the two domains is given by $f(z)$ as constructed in Fig.~\ref{fig:herschrefl}(d).

We now move to consideration of the boundary value problem on the exterior region to $B$, $\mathbb{C}\backslash B$, satisfying the same boundary conditions on $\partial B$. We proceed to show the conformal equivalence between the interior and exterior boundary value problems under the following assumptions: 1) $B$ may be constructed by sector reflections of a primitive shape and 2) the boundary conditions on $\partial B$ do not change along any primitive edge.

\subsection{The Exterior Problem: Shapes Built from Sector Reflections have \(S_i=S_e\)}\label{extprob}
\begin{figure}[tb]
    \centering
    \includegraphics*[width=\columnwidth]{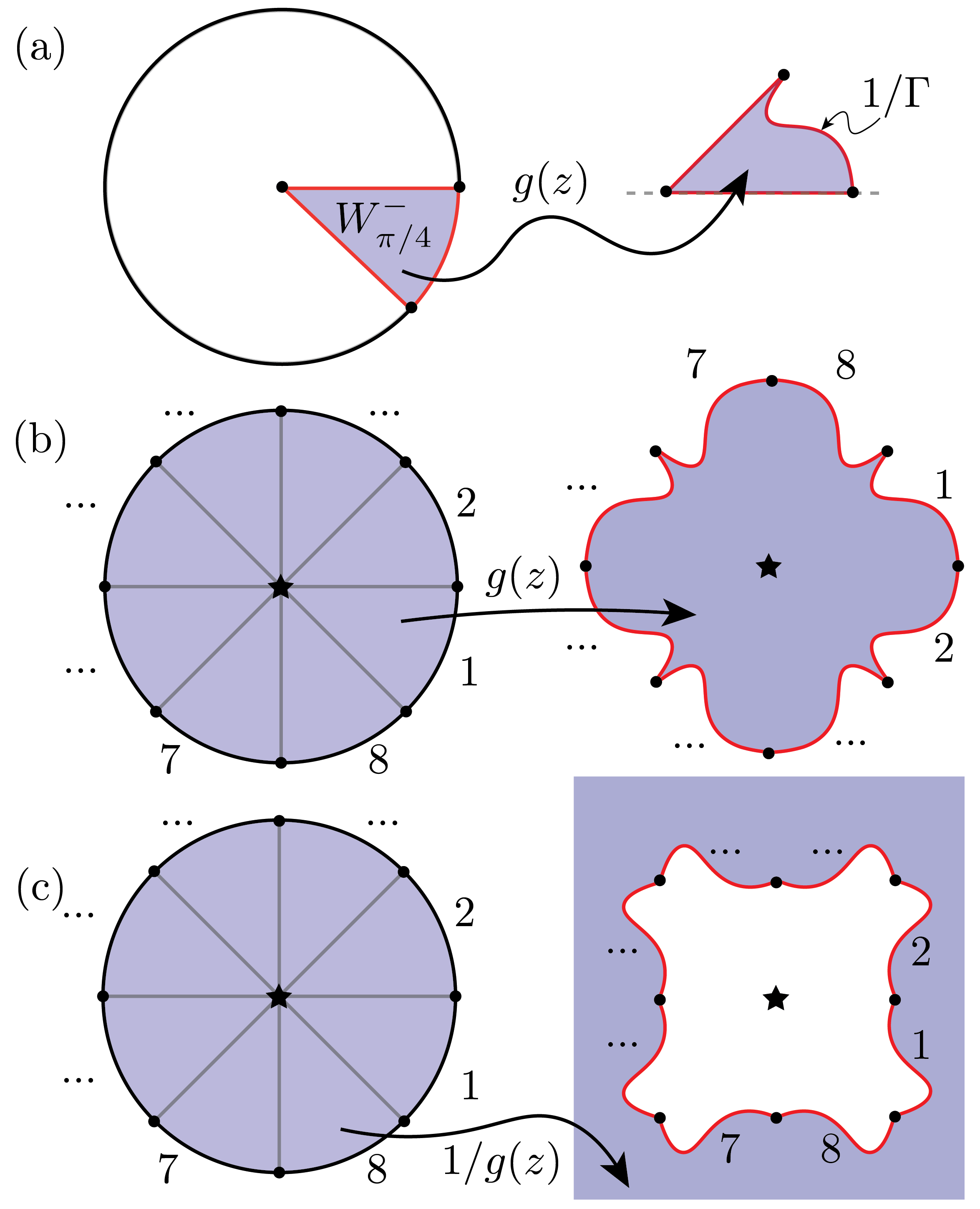}
    \caption{Construction of the map between the unit circle interior and the exterior problem, using sector reflections. The Riemann mapping theorem guarantees a mapping $g(z) $between $W_{\pi/4}^-$ and the region enclosed by $1/\Gamma$, as depicted in (a). By applying sector reflections (as was done in Fig. \ref{fig:herschrefl}) the domain of analyticity of $g(z)$ is extended to the entire unit disc as illustrated in panel (b). Note that the map in panel (b) is to an entirely new shape. The map $1/g(z)$ then sends the interior of the unit circle to exterior domain of interest.}
    \label{fig:hersch2}
\end{figure}

Consider again a Jordan curve $\partial B$ which may be constructed by $N-1$ sector reflections of a primitive edge. For illustration, let this be the same boundary curve $\partial B$ as considered in the previous section (Fig.~\ref{fig:herschrefl}(d)). We now examine the heat transfer problem exterior to that curve, in the region $\mathrm{ext}\left(\partial B\right)$, 
which is shaded on the right of Fig.~\ref{fig:hersch2}(c).
Since the shape factor is invariant under conformal mapping, we may establish the equality of the interior and exterior shape factors, relative to the curve $\partial B$, by showing that both the exterior and interior problems can be mapped onto the same problem on the interior of the unit disk (illustrated in \ref{fig:herschrefl}(d)). We now proceed to demonstrate this fact.

We now construct the conformal map from the unit circle interior to $\mathrm{ext}\left(\partial B\right)$, as outlined in Fig.~\ref{fig:hersch2}. First, consider the curve labelled $1/\Gamma$ on the right of Fig.~\ref{fig:hersch2}(a). This curve is constructed by transforming each point in the primitive curve $\Gamma$, from Fig.~\ref{fig:herschrefl}(a), according to the mapping $z\rightarrow 1/z$; the resultant primitive curve is denoted $1/\Gamma$. By applying the arguments of \S \ref{intprob} to this new primitive curve, one can construct a mapping $g(z)$ from the unit circle interior to the interior of the shape in the right side of Fig.~\ref{fig:hersch2}(b), which is itself constructed through sector reflections of $1/\Gamma$. The boundary correspondence under the mapping $g(z)$ is labelled. Note that the boundary correspondence in Fig.~\ref{fig:hersch2}(b) is now clockwise (compared to counter-clockwise in Fig.~\ref{fig:herschrefl}(d)) because here the Schwarz reflection principle was applied to the sector $1/\Gamma$ in Fig.~\ref{fig:hersch2}(a) which was \textit{above} the real axis. In a final step, the reciprocal map $1/g(z)$ sends the interior of the unit circle to $\mathrm{ext}\left(\partial B\right)$, as illustrated in Fig.~\ref{fig:hersch2}(c). The boundary correspondence is labelled, the reciprocal map having reinstated the clockwise order.

As before, if boundary conditions are unchanging over each primitive edge, the exterior shape factor problem is conformally equivalent to the problem interior to the unit circle having the same boundary conditions on equiangular arcs. Figures~\ref{fig:herschrefl}(d) and~\ref{fig:hersch2}(c) thus reveal that, under the same boundary conditions on $\partial B$, the interior and exterior shape factor problems may both be mapped to the same problem on the unit circle interior. Thus, the interior and exterior shape factor problems are conformally equivalent and possess the same shape factor and $S_e=S_i$.

Note that one may apply similar arguments to construct a mapping directly between the interior and exterior problems. 
However, we find it instructive to introduce the auxiliary problem on the unit circle interior and to show the conformal equivalence there. It should be clear to the reader that the map $1/g(f^{-1}(z))$ directly maps the interior problem (right side of Fig.~\ref{fig:herschrefl}(d)) to exterior problem (right side of Fig.~\ref{fig:hersch2}(c)) while preserving the locations of the primitive boundary segments.

\subsection{A Precise Statement of the Theorem}\label{thmhere}
The preceding sections outline a set of criteria whose fulfillment implies the equality of interior and exterior shape factors. At this stage, the more mathematical reader will beg the authors to state these criteria concisely in the form of a theorem. Consider first a definition.

Consider a partition of the complex plane, $\mathbb{C}$, into some positive whole number, $N$, of equiangular wedge sectors centered on the origin. For example, the case of $N=8$ is depicted in the left of Fig.~\ref{fig:herschrefl}(d), provided each grey line extends to complex infinity ($\abs{z}\to\infty$). 
We call this arrangement an $N$-\emph{sector partition} of the complex plane.
\begin{theorem}\label{thm:t1}\textbf{(Interior/Exterior Shape Factor Equality)}
Consider an $N$-sector partition of the complex plane, for an even value $N \geq 4$.
Now consider a Jordan curve $\partial B$ constructed from $N-1$ sector reflections of a primitive edge.
Suppose that each primitive edge  of $\partial B$ is subjected to either a zero Neumann (adiabatic) or a constant Dirichlet (isothermal) boundary condition, with the Dirichlet constant restricted to one of two temperatures. Additionally, Dirichlet sections at different temperatures must be separated by at least one adiabatic section. Then, the shape factors for heat transfer in the domains $\mathrm{int}\left(\partial B\right)$ and $\mathrm{ext}\left(\partial B\right)$ are equal.
\end{theorem}

Note that we require $N$ to be greater or equal to four because the boundary conditions may not change along any primitive edge. Since the simplest heat transfer problem having a finite shape factor comprises two isothermal segments separated by two adiabatic segments, a minimum of $N=4$ is required to specify all four segments in accordance with the boundary condition criterion.

\section{The Unit Disc, Regular \(K\)-gons, and Beyond}
We now explain, using the theoretical tools developed in \S \ref{intprob} and \S \ref{extprob}, why the boundary value problems analysed by Lienhard~\cite{lienhard2019exterior} happened to have $S_i=S_e$. We shall show that all cases considered therein were within the scope of theorem \ref{thm:t1}.  We also discuss more general shapes and Yin-Yang bodies.

\subsection{The Square}\label{sq}
\begin{figure}[!t]
    \centering
    \includegraphics*[width=\columnwidth]{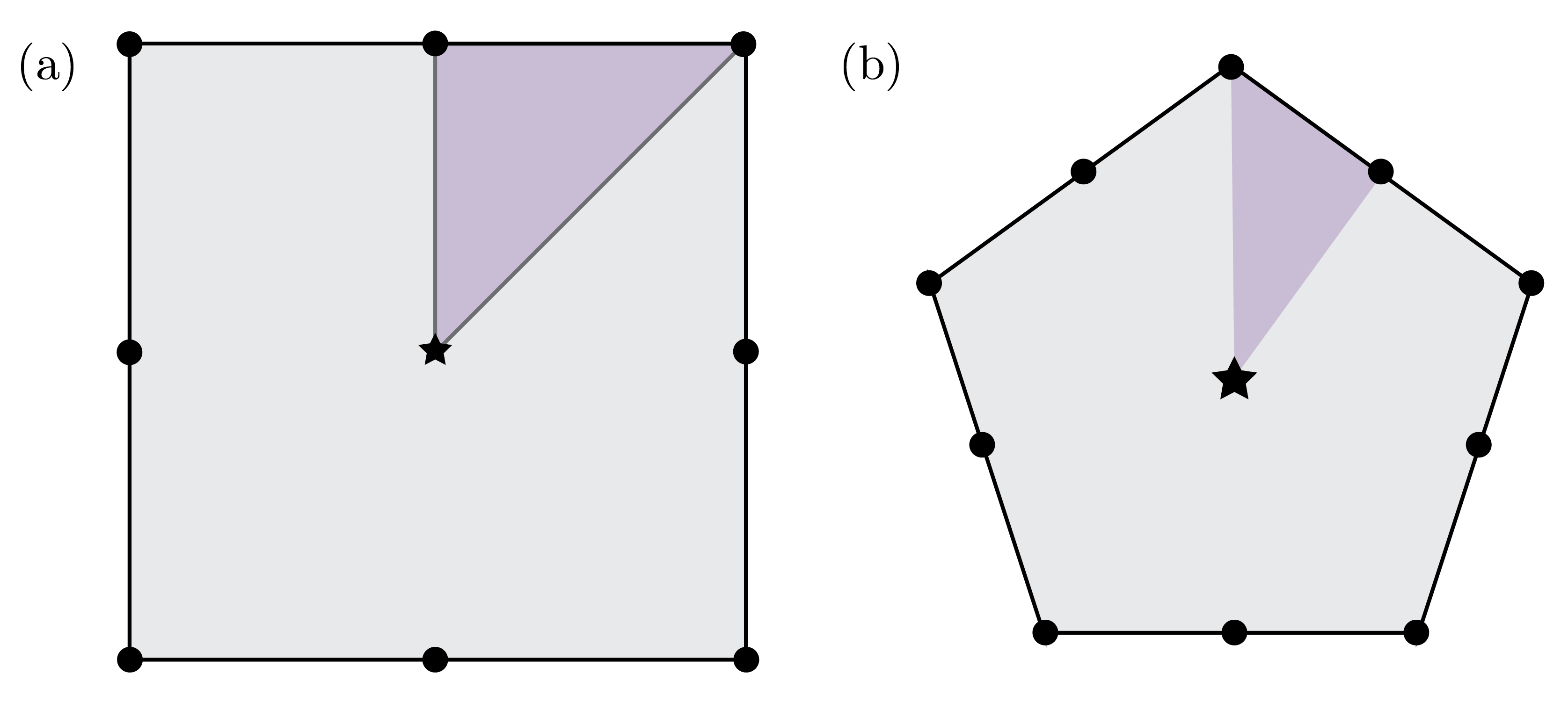}
    \caption{
    Example primitive sectors for the (a) square ($N=8$) and (b) pentagon ($N=10$) are shaded in purple. The conformal map from the interior of the unit disk to the square maps each wedge spanning $2\pi/N$ to each of the primitive shapes in Fig.~\ref{fig:primitives}. }
    \label{fig:primitives}
\end{figure}

The square interior domain comprises eight of the primitive triangles shaded in purple in Fig.~\ref{fig:primitives}(a). It follows from the developments of \S \ref{intprob} and \ref{extprob} that the mapping from the square's interior to its exterior will preserve the eight primitive edges (segments connecting each pair of adjacent black circles in Fig.~\ref{fig:primitives}(a)). 
Lienhard~\cite{lienhard2019exterior} numerically tested two configurations of boundary conditions on the square. Both examples considered therein specified boundary conditions unchanging along each primitive edge, in accordance with the assumptions of Theorem~\ref{thm:t1}, and it was indeed found that $S_{i}=S_{e}$.

 Nonetheless, the equality of interior and exterior shape factors does not generalize to arbitrary problems. Symmetries of both the domain boundary shape and the boundary condition specification are required to guarantee the equality of interior and exterior shape factors: 1) the shape interior $\mathrm{int}\left(\partial B\right)$ must be constructable from sector reflections of a primitive shape; and 2) boundary conditions must be unchanging along each primitive edge.

\subsection{Regular \(K\)-gons}\label{kgons}

A regular $K$-gon can be constructed from $N=2K$ copies of a primitive triangle; we illustrate the primitive triangles for the cases $K=4$ and $K=5$ in Fig.~\ref{fig:primitives}. If only one boundary condition is specified as unchanging along each primitive edge, then the interior and exterior shape factors are guaranteed to be equal, in accordance with Theorem~\ref{thm:t1}.

Specifically for regular $K$-gons, primitive edges are half-faces. Thus, a single boundary condition must be specified along the entirety of each half-face to ensure that $S_i=S_e$. For example, the square of \S \ref{sq} has $K=4$, but boundary conditions are allowed to vary between its $N = 2K= 8$ primitive edges (half-faces), for Theorem~\ref{thm:t1} to be applicable and thus to ensure the equality of interior and exterior shape factors.

To validate this somewhat surprising result, we use FEM to compute the interior and exterior shape factor of the hexagon with half-heated faces in the last row of Table~\ref{tab:FEM-exact}. Here, the top and bottom faces of the hexagon have opposite halves at different temperatures, while the remainder of the hexagon boundary is taken to be adiabatic. The computed interior and exterior shape factors are shown together with the analytical result for the interior problem given by Hersch \cite{hersch1982harmonic}, $S_{i}=1/\sqrt{3}$. The interior value is within 0.03\% of the exact result and the exterior value is within 1.3\%. While the exact solution $S_{i}=1/\sqrt{3}$ given by Hersch was computed for the interior problem, Theorem~\ref{thm:t1} indicates the same exact result for the exterior problem, as is supported by our numerical calculations.

\subsection{The Disc}
For the disc, the map $1/z$ takes the interior problem to the exterior problem with the same boundary conditions (up to a reflection, which does not affect the shape factor). Thus, for any alternating sequence of constant Dirichlet and Neumann conditions around the boundary of a circle, the interior and exterior problems are conformally equivalent and $S_i=S_e$ must hold.

In light of \S \ref{kgons}, one may also rationalize the equality of the interior and exterior shape factors of the circle in the following manner. Consider the circle as the limit of a regular $K$-gon, whose vertices lie on the unit circle, as $K\rightarrow\infty$. For the $K$-gon, the interior and exterior shape factors are guaranteed to be equal as long as boundary conditions are constant along each of the $2K$ primitive edges. In the limit of $K\rightarrow \infty$, the length of each primitive edge tends to zero. In this limit, any finite segment of the circle can be decomposed into a union of primitive segments. Thus, for any shape factor problem on the circle, with finite boundary segments possessing different boundary conditions, it is guaranteed that $S_e=S_i$.

\subsection{Other Shapes}
\begin{figure}[!t]
    \centering
    \includegraphics*[width=\columnwidth]{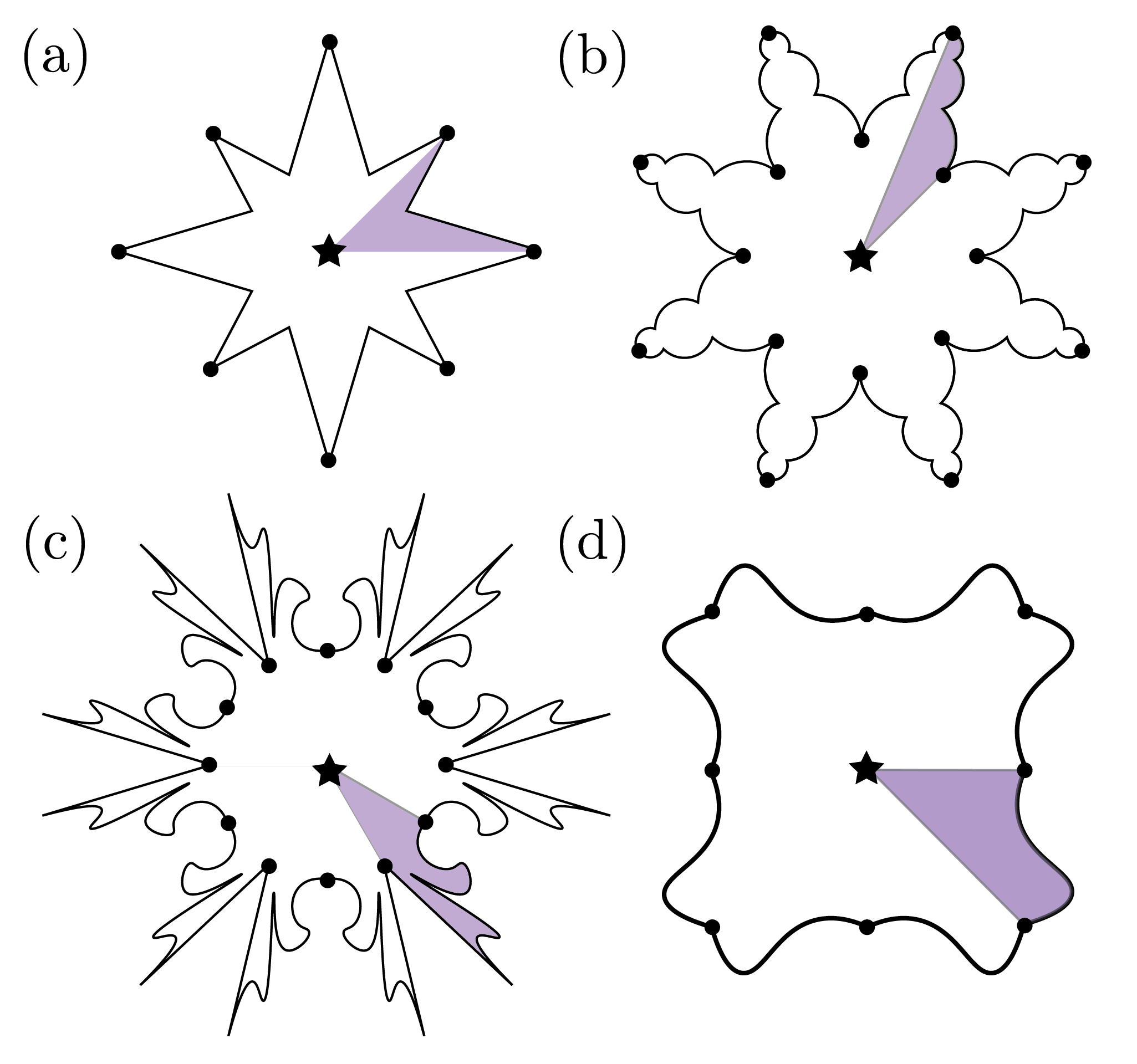}
    \caption{
    Example of some irregular shapes that may be constructed from sector reflections of a primitive shape (shaded in purple) for cases of (a) $N=8$, (b) $N=16$, (c) $N=12$, and (d) $N=8$.}
    \label{fig:symmexamps}
\end{figure}

Note that our criteria, which guaranteed $S_i=S_e$, do not require shapes to be regular polygons---or any other simple shape for that matter. The first criterion (geometric criterion) is that the domain may be constructed from sector reflections of a primitive curve confined to a sector of angle $2\pi/N$ with $N \geq 4$ being some even integer. Some examples of more complex objects are given in Fig.~\ref{fig:symmexamps}, with primitive shapes shaded in purple. Note that for Theorem~\ref{thm:t1} to apply, ensuring the equality of interior and exterior shape factors, applied boundary conditions must be unchanging along each primitive edge.

\subsection{Yin-Yang Bodies have \(S_i=S_e=1\)}\label{sec:yiny}
We here demonstrate how Schwarz reflections may be used to prove another interesting symmetry of shape factors, the Yin-Yang property~\cite{lienhard1981heat}, which is outlined as follows.

\begin{figure}[!t]
    \centering
    \includegraphics*[width=\columnwidth]{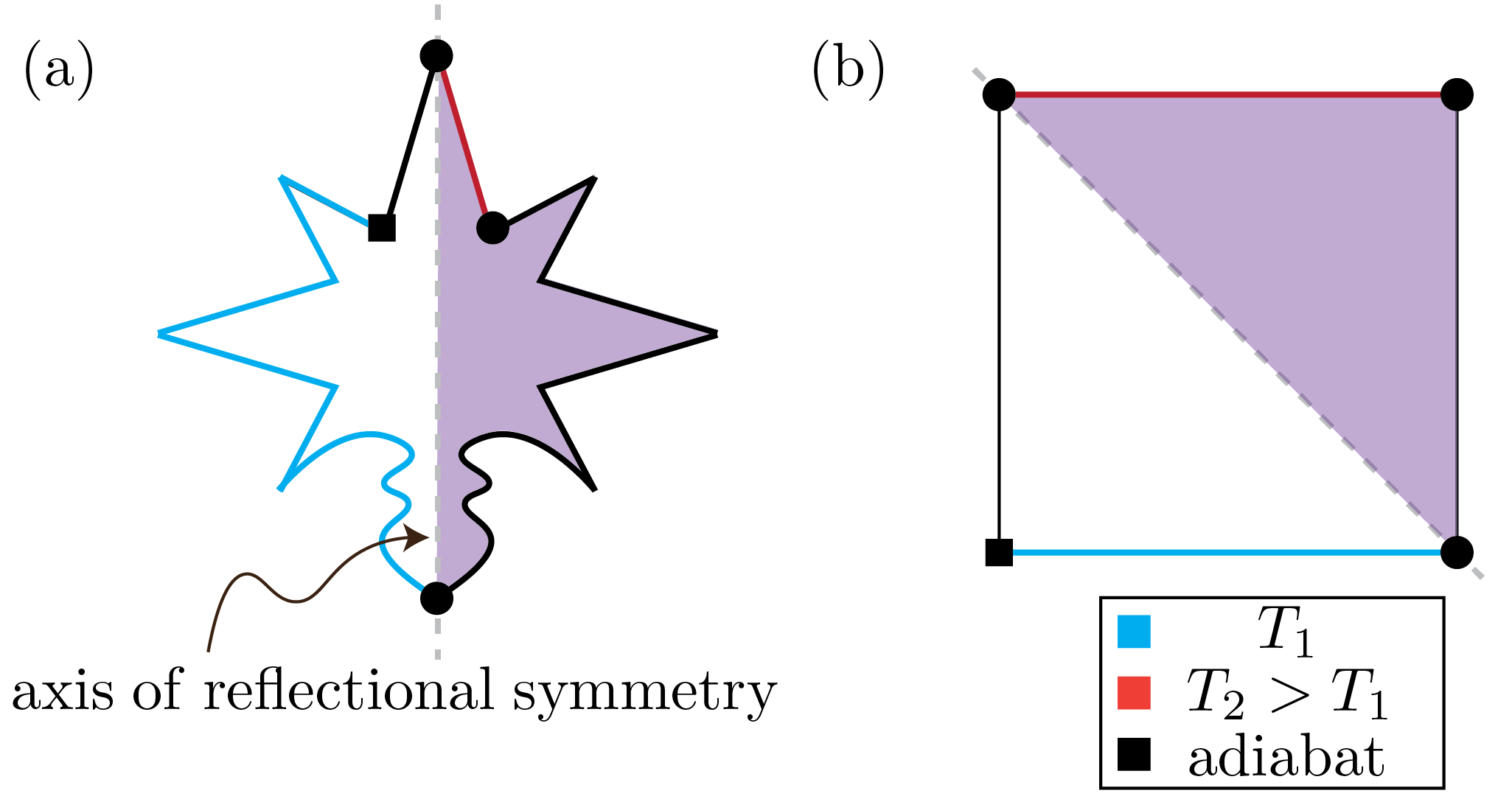}
    \caption{
    Conformal equivalence of Yin-Yangs and the square problem with $S=1$. The domain in panel (a) embodies the Yin-Yang symmetry as described in \S \ref{sec:yiny}. By the Riemann mapping theorem, there exists a mapping between the purple shaded section of the domain in (a) and the half-square with the drawn boundary correspondence as drawn in panel (b).The Schwarz reflection principle then gives the remainder of the map (white region with the illustrated boundary correspondence. Since the shape factor of~(b) is trivially equal to unity, by the conformal invariance of the shape factor, $S=1$ in~(a) as well.
    }
    \label{fig:yinfig}
\end{figure}

Suppose a Jordan curve $\partial B$ has an axis of reflectional symmetry, as shown in Fig.~\ref{fig:yinfig}. The boundary on one side of the axis contains an isothermal segment and an adiabatic segment; on the reflected side, the isothermal and adiabatic conditions have been interchanged.
When a shape factor problem embodies all the aforementioned symmetries, we say it embodies the Yin-Yang symmetry. 
Lienhard~\cite{lienhard1981heat} stated that Yin-Yang problems have a shape factor of unity, although no formal proof was given.  

Here, we use Schwarz reflections to provide a proof that all such shape factor problems are conformally equivalent to a problem on the square which possesses $S=1$. Consider a shape factor problem embodying the Yin-Yang symmetry as in Fig.~\ref{fig:yinfig}(a). Now consider single half of the domain (shaded purple), as defined by a cut made along the axis of symmetry (grey dotted line). By the Riemann mapping theorem, there exists a mapping between the purple region in Fig.~\ref{fig:yinfig}(a) and the purple half of the square (Fig.~\ref{fig:yinfig}(b)), wherein we are free to dictate the image location of three boundary points.  In particular, we choose to specify the images of the three following boundary points (shown with black circles): the two intersections of $\partial B$ with the reflection axis, and the point where the isothermal boundary segment adjoins the adiabatic boundary segment. We choose as the images of these three points the three vertices of the square also indicated by black circles. The boundary condition correspondence is coloured in both panels. The Schwarz reflection principle then implies the image of the remainder of the domain (white region), for which the boundary correspondence is also indicated in Fig.~\ref{fig:yinfig}. Since the shape factor of Fig.~\ref{fig:yinfig}(b) is trivially $S = 1$, the conformal invariance of the shape factor shows that $S=1$ in the Fig.~\ref{fig:yinfig}(a) as well. 

By a similar argument to \S \ref{extprob}, the proof also holds for the exterior Yin-Yang problem, so that the interior and exterior Yin-Yang shape factor problems must have $S_i=S_e=1$.

Finally, we note that the Yin-Yang problem shows that, although our criteria in Theorem~\ref{thm:t1} are sufficient to guarantee $S_i=S_e$, they are not necessary criteria:  many Yin-Yangs do not meet the criteria of Theorem~\ref{thm:t1}, but all have equality of their interior and exterior shape factors.

\newcommand\AddFig[2]{\raisebox{-0.5\height}{\includegraphics[keepaspectratio=true,height=#1\columnwidth]{figures/#2}}}
\begin{table*}[!t] 
\caption{Shape factors for selected \(K\)-gons, comparing known exact values to interior and exterior FEM results. Far field boundary conditions for the exterior are: square---dipole temperature field with radius~20; hexagon---adiabatic outer boundary with radius~30; octagon---adiabatic outer boundary with radius~35. Each polygon has sides of length~2. Interior and exterior flux plots with isotherms in black and adiabats in red, except second and third exterior cases, which show temperature field and isotherms. Note that the fifth case satisfies $N = 2K = 12$, as discussed in \S \ref{kgons}, with boundary conditions specified on half-sides.}\label{tab:FEM-exact}
\centering
\begin{tabular}{c@{\hspace*{2em}}ccccc}
\toprule
$\bm{N}$ & \textbf{Interior} & $\bm{{S_i}}$ \textbf{(FEM)} & $\bm{S}_{\textbf{exact}}$ & $\bm{{S_e}}$ \textbf{(FEM)} &  \textbf{Exterior} \\
\midrule
$8$ & \AddFig{0.38}{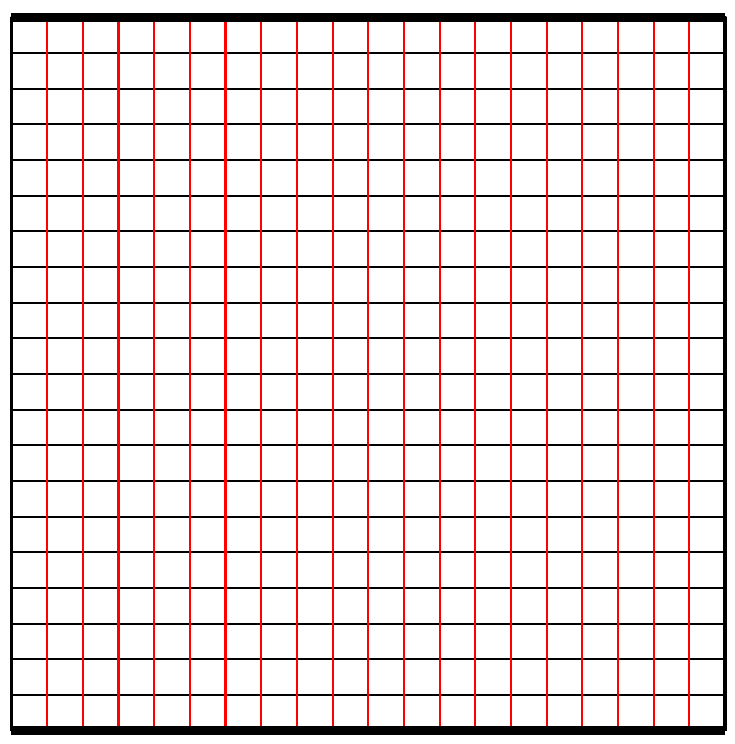} & 1.0000 & 1 & 0.9948 & \AddFig{0.4}{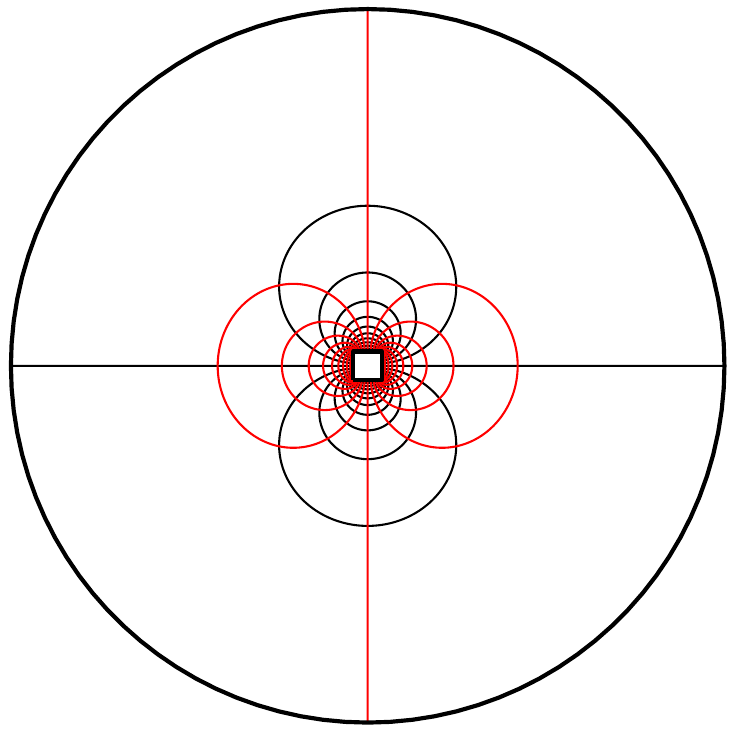} \\[6em]
$12$ & \AddFig{0.4}{hex-S21.pdf}    & 0.9998 & 1 & 0.9882 & \AddFig{0.4}{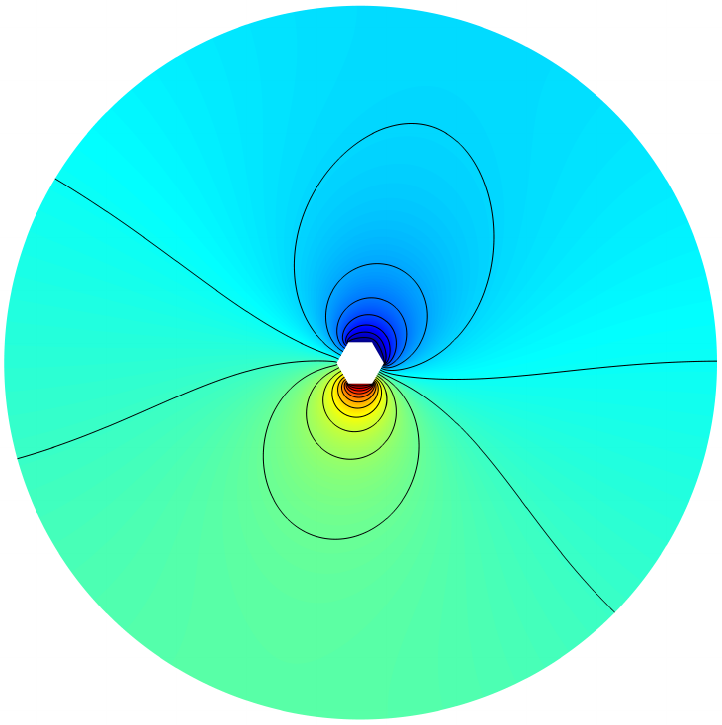} \\[6em]
$12$ & \AddFig{0.4}{hex-S2n1.pdf}
 & 1.1544  & \parbox[t]{4.8em}{\centering $1.15470\cdots$ \rule{0pt}{15pt}$\big(2\big/\sqrt{3}\big)$ }  & 1.1427 & \AddFig{0.405}{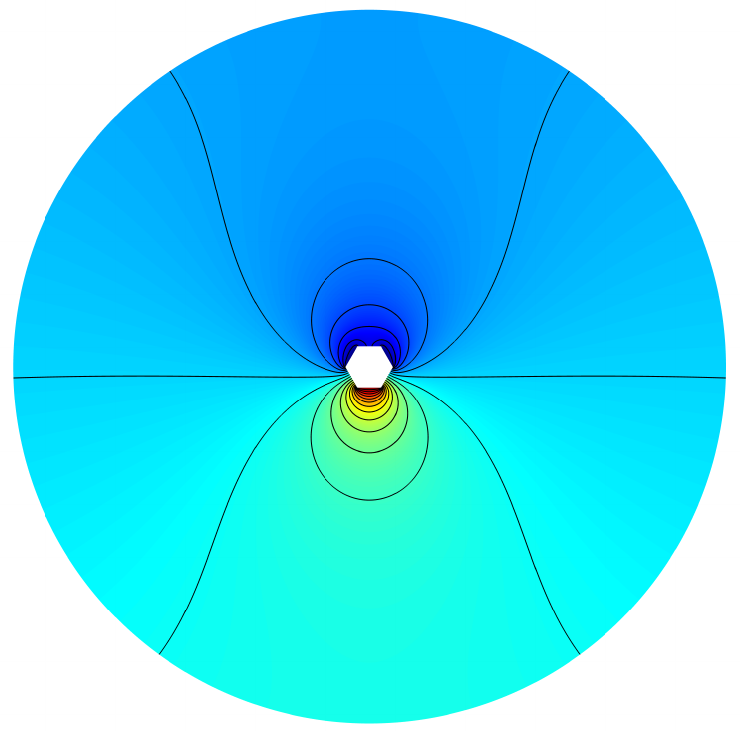} \\[6em]
$16$ & \AddFig{0.4}{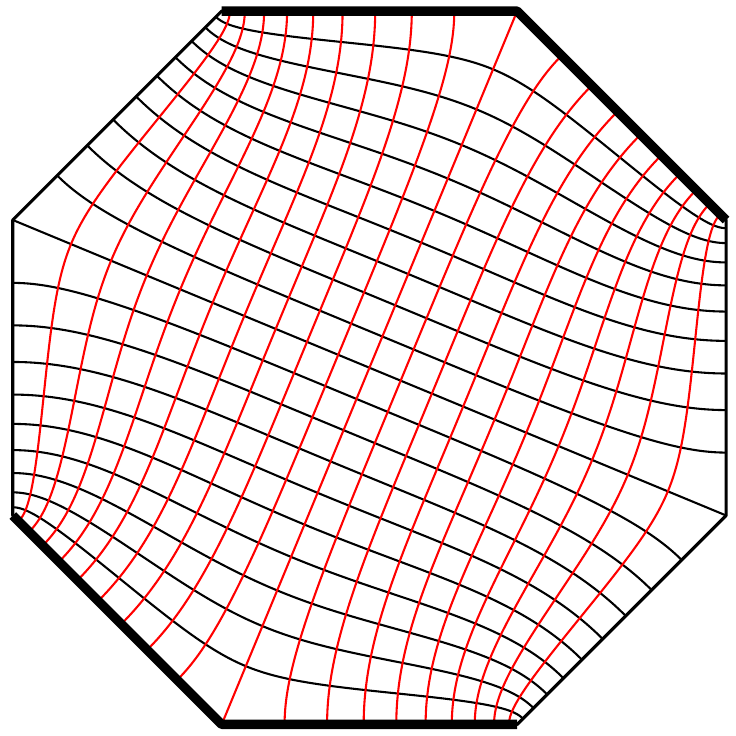}   & 0.9994 & 1 & 0.9896 & \AddFig{0.4}{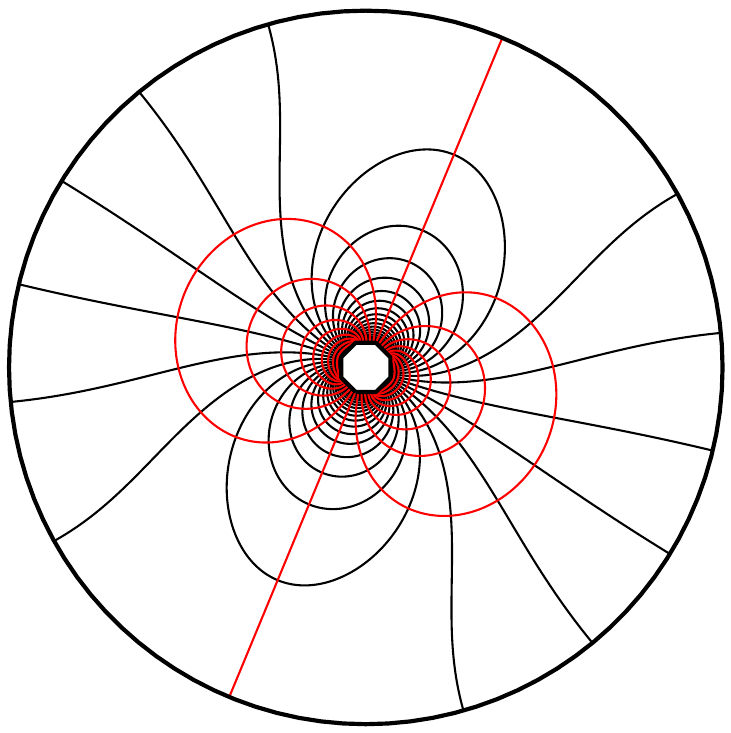} \\[6em]
$12$ & \AddFig{0.4}{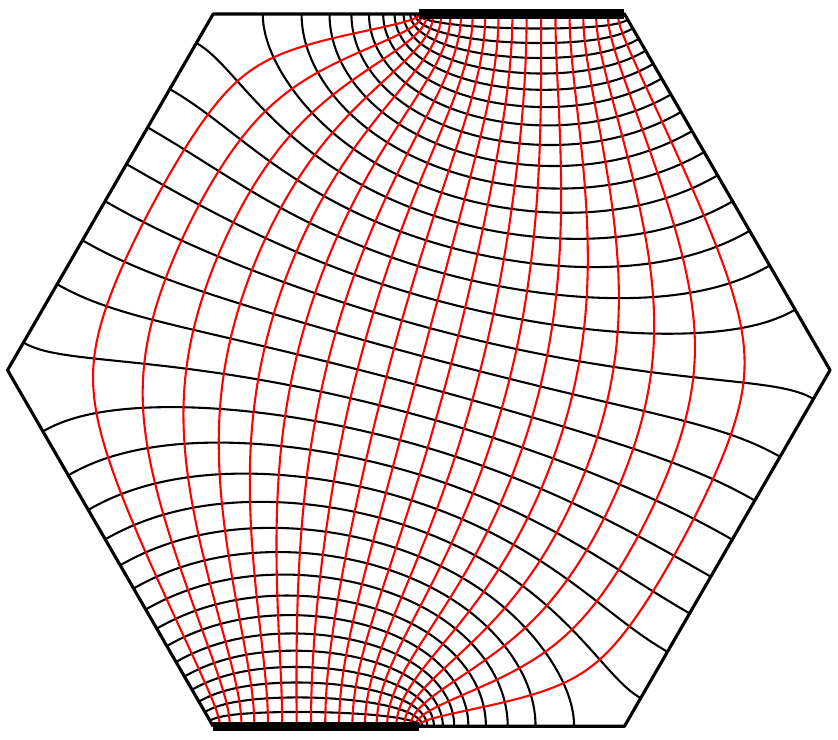}   & 0.5776 & \parbox[t]{4.8em}{\centering $0.57735\cdots$ \rule{0pt}{15pt}$\big(1\big/\sqrt{3}\big)$}  & 0.5700 & \AddFig{0.4}{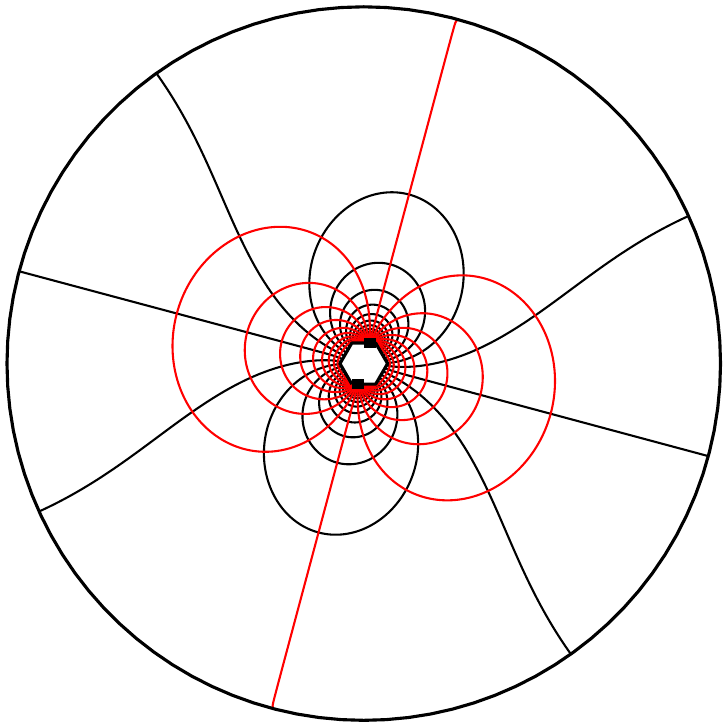} \\[6em]
\bottomrule
\end{tabular}
\end{table*}

\section{Validation for Several Cases}
We now show several explicit results to support our conclusion.

\subsection{Interior and Exterior Numerical Results Compared to Exact Shape Factors}
Table~\ref{tab:FEM-exact} contains several cases for which exact interior shape factors are known. Interior and exterior simulations are shown, along with the computed shape factors.  The interior FEM results are accurate to 0.05\% or better (see Appendix~\ref{sec:FEM-conv}). The exterior FEM results are accurate to about 1.5\%, with most of the error attributable to a finite outer domain.  In all cases, the exact, interior, and exterior results agree to within the apparent accuracy of the FEM computations.  Additional numerically validated $K$-gons are shown in Appendix~\ref{sec:catalog}.

Note that in the third row in Table~\ref{tab:FEM-exact} the two upper segments have the same temperature, but are not connected. Theorem~\ref{thm:t1} applies to such cases without limitation.

The exact values for the first, second, and fourth cases in Table~\ref{tab:FEM-exact}, $S = 1$, arise because these problems have Yin-Yang symmetry~(\S\ref{sec:yiny}). The exact results for third and fifth cases were found by Hersch for the interior geometries ~\cite{hersch1982harmonic}. We have proven herein that Hersch's exact values for the interior domain are also those for the exterior problem. Thus, we label these exact results $S_{\mathrm{exact}}$ since there is distinction between the interior and exterior shape factors. Our numerical results support our proof.

\subsection{The Compass Rose}
\renewcommand\AddFig[2]{\raisebox{-0.5\height}{\includegraphics[keepaspectratio=true,height=#1\columnwidth]{figures/#2}}}
\begin{table*}
\caption{Shape factors for the Compass Rose. The temperature field and isotherms are shown. The far field boundary is adiabatic with radius~30. The long arms have a radial length of 1.5, the short arms have 1.17, and the inner vertices have 0.6.}\label{tab:compass-rose}
\centering
\begin{tabular}{c@{\hspace*{3em}}ccccc}
\toprule
$\bm{N}$ & \textbf{Interior} & $\bm{{S_i}}$ \textbf{(FEM)} & $\bm{S}_{\textbf{exact}}$ & $\bm{{S_e}}$ \textbf{(FEM)} &  \textbf{Exterior (near field)} \\
\midrule
\rule{0pt}{6.5em}$4$ & \AddFig{0.45}{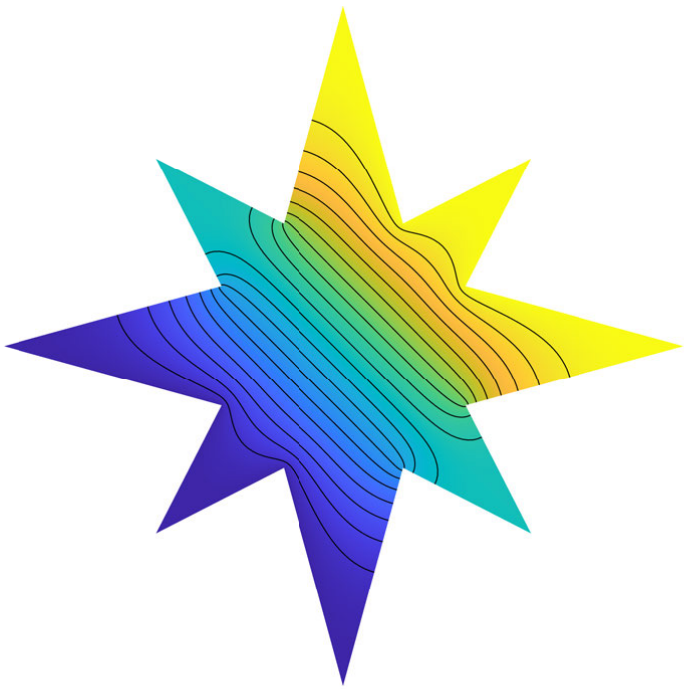}  & 0.9993 & 1  & 0.9749 & \AddFig{0.45}{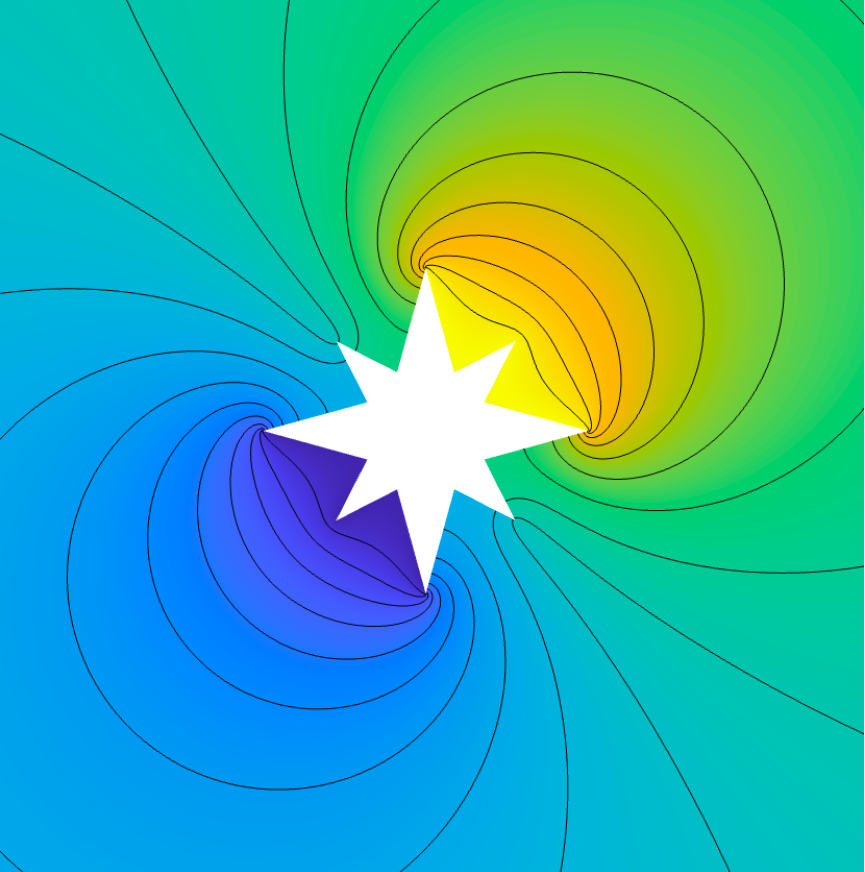} \\[7em]
$8$ & \AddFig{0.45}{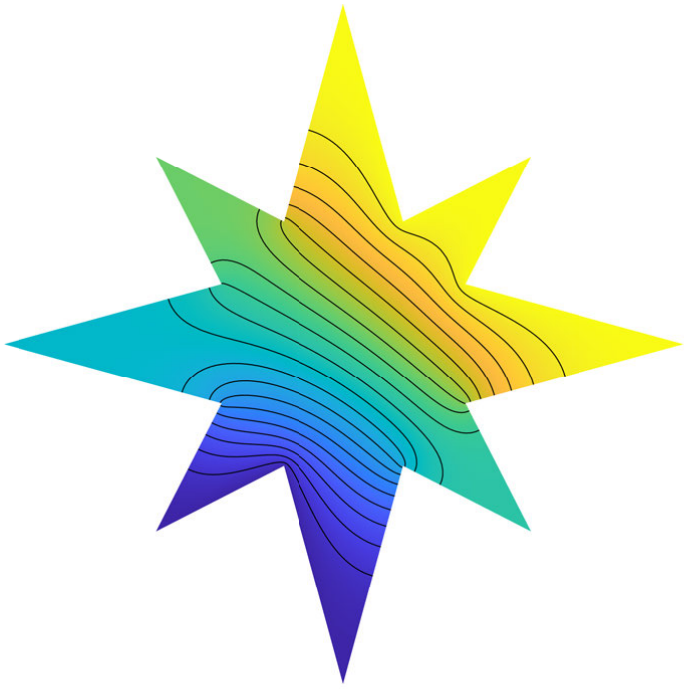} & 0.8191 & n/a & 0.7907 & \AddFig{0.45}{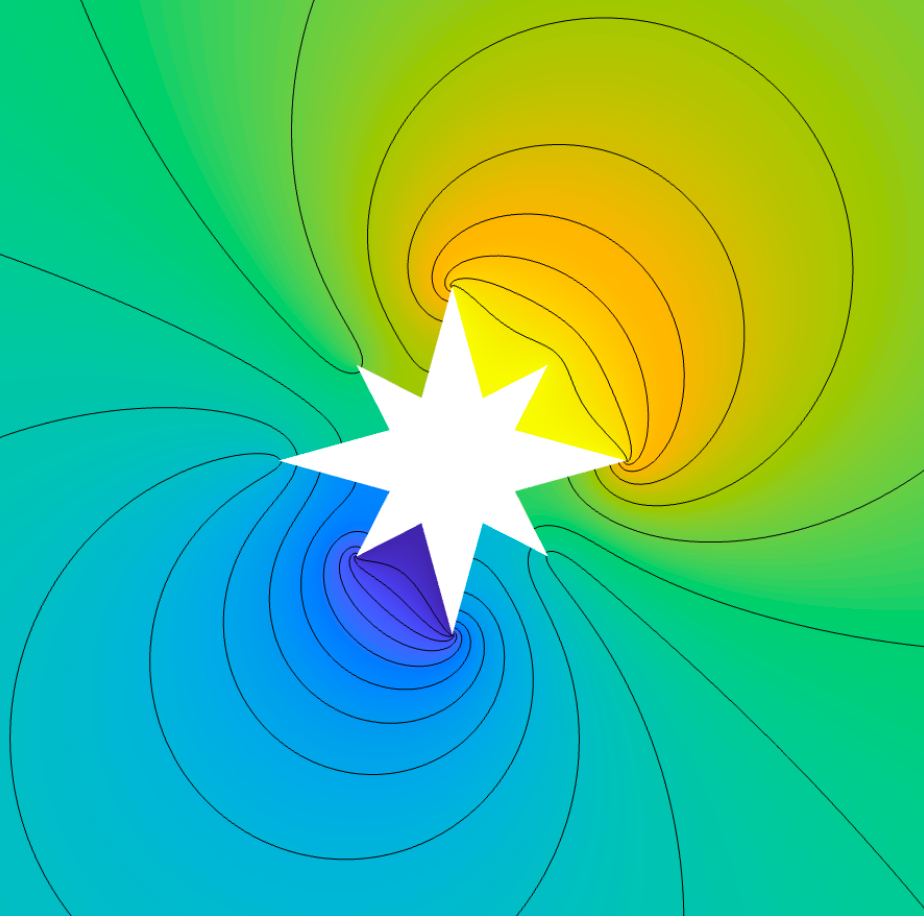} \\[6.5em] 
\bottomrule
\end{tabular}
\end{table*}

The Compass Rose is a design traditionally used to display the cardinal directions on a map, consisting of four orthogonal arms of one length and four of another at a 45\textdegree\ rotation. In the language of the present paper, the Compass Rose has an $N=8$ symmetry with a primitive element containing one-half arm of each length (Fig.~\ref{fig:primitives}(a)).

To further demonstrate the applicability of our theorem, we have simulated heat conduction inside and outside the rose for two arrangements of boundary conditions compatible with Theorem~\ref{thm:t1} (see Table~\ref{tab:compass-rose}). The first case specifies isothermal boundary conditions on four identical sections of the boundary, with edges in the upper right-hand and lower left-hand quadrants at different temperatures. This case embodies the Yin-Yang symmetry so that $S_{\textrm{exact}} = 1$. 

The second case has isothermal boundary conditions specified in an asymmetric fashion. The two primitive elements in the upper right-hand quadrant have edges at one temperature, and one primitive element in the lower right-hand quadrant a different temperature. 

The numerical value of $S_i$ for the first case is within $0.07\%$ of the theoretical value. The numerical value of $S_e$ is $2.5\%$ below the theoretical value, primarily owing to the difficultly in capturing the very strong singularities of the heat flux at the tips of the long arms. A further difficulty with this, and other, exterior computations is the need to truncate the infinite domain with an outer boundary of sufficient size (see Appendix~\ref{sec:FEM-kgon}). If nothing else, these difficulties illustrate the great value in knowing that $S_e$ can be found through the much easier computation of $S_i$.
We have no theoretical value for the second case; the numerical values differ by 3.5\%. 

\subsection{Known Conformal Maps for \(K\)-gons}
The explicit Schwarz-Christoffel transformations between the unit circle and the $K$-gon interior and exterior, show that the interior and exterior problems are conformally equivalent when each $K$-gon face adopts a single boundary condition~\cite{nehari}.
These transforms are discussed briefly in Appendix~\ref{sec:SC-maps}.
While the Schwarz-Christoffel maps show that $S_e=S_i$ when boundary conditions do not vary over each full face, our results in \S \ref{extprob} indicate a less restrictive requirement: boundary conditions must not vary over each \textit{half}-face.


\section{Analogues in Darcy Flows and Electrostatic Capacitance }\label{analogies}
The results presented in \S \ref{sec:syms} have precise mathematical analogues in fluid flow through porous media and in electrical capacitance, which we now briefly outline. 

\subsection{Darcy Flow Through Porous Media}
Pressure-driven flow through a porous media has a velocity given by $\boldsymbol{u}=-\sigma \nabla P$, where $\sigma$ is the hydraulic conductivity, and $P$ is the pressure. Mass conservation for an incompressible fluid then implies that the pressure satisfies the Laplace equation. Consider a situation where flow is driven between isobaric segments held at two different pressures, and separated by impermeable segments, with a pressure difference $\Delta P$. Then, the volume flux between the isobaric segments, $Q_f$, can be written in terms of a flow shape factor $S_f$ which depends on the geometry according to
\begin{equation}
    Q_f=S_f \sigma \Delta P.
\end{equation}

The present results apply to flow through a porous media after the following modifications: $k$ is replaced by the hydraulic conductivity $\sigma$; $\Delta T$ is replaced by the pressure differential between isobaric segments $\Delta P$; the heat flux $Q$ is replaced by the volume flux $Q_f$; and the shape factor is replaced a flow shape factor $S_f$. Note also that in the fluid flow context, thermal conductors and insulators are replaced by isobaric segments and impermeable boundaries, respectively. In the fluid flow context, Theorem~\ref{thm:t1} guarantees the equivalence of the flow shape factors interior and exterior to a body.

\subsection{Electrostatic Capacitance}
When a voltage differential is established between two perfect electrical conductors, a charge of magnitude $q$ is established on each conductor. For a given voltage differential $\Delta V$, the amount of charge $q$ is characterized by the electrostatic capacitance $C$, which is a function of the geometry, according to 
\begin{equation}
    q=C \Delta V.
\end{equation}
Our results apply to the electrical capacitance, $C$, after the following substitutions: $\Delta T$ is replaced by electrical potential difference between conducting segments $\Delta V$; $Q/k$ is replaced by the conductor surface charge $q$; and the shape factor is replaced by the electrical capacitance $C$.
Thermal conductors and insulators must be replaced by electrical conductors and insulators, respectively. Theorem~\ref{thm:t1} guarantees the equivalence of the electrostatic capacitances interior and exterior to a body.


\section{Summary}
Interior and exterior shape factors are not universally equal, contrary to a previous report. In this paper, we first demonstrated why the interior and exterior shape factors are not always equal. We then provided a set of criteria sufficient to guarantee equality.  In particular, the interior and exterior shape factors \textit{will be equal} if specific symmetries are present: 1) the shape boundary must be constructed from sector reflections of a primitive sector edge; and 2) the boundary conditions — either zero Neumann (adiabatic) or constant Dirichlet (isothermal) — must not change along each primitive edge.  This result is stated precisely in Theorem~\ref{thm:t1}.  Objects that meet the conditions of the theorem include not only regular polygons and discs, but also much more complex shapes like those shown in Fig.~\ref{fig:symmexamps}. These results have direct analogies in flow through porous media and in electrostatic capacitance, as outlined in~\S \ref{analogies}.

We have tested our theorem against independent results from finite-element simulations and from Schwarz-Christoffel maps. Additionally, a brief tabulation of shape factors in geometries that meet the conditions of Theorem~\ref{thm:t1} is provided in Appendix~\ref{sec:catalog}.

We note that this paper may be useful also as an accessible tutorial for applying symmetry methods to Laplace problems through the exploitation of the Schwarz reflection principle.

\section*{Acknowledgment} 
K.M. would like to thank Alex Cohen and Darren Crowdy for valuable discussions. Both authors would like to thank Nick Trefethen for useful comments and for pointing out the work of Hersch.

\section*{Funding Data}
K.M. was supported by a MathWorks Fellowship during this work. He was also supported by a Chateaubriand Fellowship and hosted at ESPCI, Paris during the completion of the paper.


\begin{nomenclature}
\entry{$B$}{a planar body (interior of a Jordan curve, $\partial B$)}
\entry{$a,b$}{dimensions of rectangle}
\entry{$b_k$}{pre-vertices on unit disc}
\entry{$c_1,c_2,c_3$}{constants, App.~\ref{sec:SC-maps}}
\entry{$\mathbb{C}$}{the set of complex numbers}
\entry{$\mathrm{ext}(\partial B)$}{region exterior to~$\partial B$}
\entry{$f(z), g(z)$}{conformal maps}
\entry{$H_{\textrm{max}}$}{maximum mesh size for FEM}
\entry{$\mathrm{int}(\partial B)$}{region interior to~$\partial B$}
\entry{$k$}{thermal conductivity (W m$^{-1}$ K$^{-1}$)}
\entry{$K$}{number of sides in polygon}
\entry{$N$}{number of primitive elements composing a shape}
\entry{$Q$}{heat flow (W m$^{-1}$)}
\entry{$S$}{shape factor (--)}
\entry{$T$}{temperature (K)}
\entry{$x$}{coordinate along the bottom wall, App.~\ref{sec:rectangles}}
\entry{$W$}{domain in the complex $z$ plane}
\entry{$z$}{coordinate in the complex plane}

\EntryHeading{Greek Letters}
\entry{$\alpha$}{angle on disc, see Fig.~\ref{fig:disc-alpha}}
\entry{$\gamma$}{outer section of boundary 
denoted a primitive edge, Fig.~\ref{fig:herschrefl}a}
\entry{$L_1$, $L_2$}{Straight lines along each sector boundary extending from the origin to the ends of the primitive edge, Fig.~\ref{fig:herschrefl}a}
\entry{$\Gamma$}{closed curve defined by $\Gamma=L_1\cup L_2 \cup \gamma$, and denoted the primitive curve}
\entry{$\Delta T$}{temperature difference (K)}
\entry{$\delta x$}{$H_{\textrm{max}}/2$}
\entry{$\eta_1$}{$K/[2(K+2)]$}
\entry{$\mu_i$}{exterior angles as fraction of $\pi$}
\entry{$\Omega$}{range in the mapped plane}

\EntryHeading{Superscripts and Subscripts}
\entry{$e$}{exterior value}
\entry{$h$}{hexagon}
\entry{$i$}{interior value}
\entry{$p$}{pentagon}
\entry{$\overline{z}$}{complex conjugate of $z$}

\EntryHeading{Acronyms}
\entry{FEM}{finite element method}

\end{nomenclature}


\appendix
\section{FEM convergence}\label{sec:FEM-conv} 
\subsection{Rectangles}\label{sec:rectangles}
For a rectangle with isothermal sides of width $a$ and adiabatic sides of height $b$, the interior shape factor is exactly $S_i = a/b$. The exterior shape factor was computed using \textsc{Matlab}'s finite-element method (FEM) tools. Mesh convergence was evaluated by reducing the maximum mesh size, $H_{\textrm{max}}$, until machine memory limits were exceeded (Table~\ref{tab:rect-conv}).  Localized mesh refinement was applied at each outside vertex and edge, limiting the maximum mesh to $H_{\textrm{max}}/40$. The computation is sensitive to the outer boundary condition of the domain, which was set to a circle of radius~20. The far field temperature was shown to limit to a dipole distribution in \cite{lienhard2023potential}, and that boundary condition was used on the outside edge.  

The heat flow was determined by integrating the computed heat flux along the bottom edge. The heat flux has an integrable singularity at each vertex ($x =\pm1$), so corner corrections were applied as described in \cite{lienhard2023potential}. Specifically, the computed heat flux normal to the bottom boundary was fit the first and third terms of the expansion for corner singularities (eqn.~(68) in \cite{lienhard2023potential}):
\begin{multline}
\abs*{\frac{\partial T}{\partial n}}  =  A \eta_1 (1+x)^{\eta_1-1} + B \eta_1 (1-x)^{\eta_1-1} \\ + C (5\eta_1)(1+x)^{5\eta_1-1} + D (5\eta_1) (1-x)^{5\eta_1-1}
\end{multline}
where $x$ is the coordinate along the bottom wall, $\eta_1 = K/[2(K+2)]$ for a regular $K$-gon or a 
rectangle ($K=4$), and $A,B,C,D$ are fitting coefficients.
When curve fitting the flux and when integrating for the shape factor $S$, both ends of the bottom edge are omitted over a distance $\delta x =H_{\textrm{max}}/2$, based on the observed break-down of the flux calculation near the endpoints.  Integrating the fitted flux distribution a distance $\delta x$ from each corner gives
\begin{equation}
 \Delta S\mkern2mu \Delta T \cong (A+B)(\delta x)^{\eta_1}  + (C+D)(\delta x)^{5\eta_1}
\end{equation}
for $k=1$. We then add $\Delta S$ to $S$.

The horizontal dimension of the rectangle was held fixed at $a=2$ and $b$ was varied. Table~\ref{tab:rect-conv} shows the results. The computed value for $a/b =1$ is within 0.5\% of the exact value, 1, when $H_{\textrm{max}} = 0.02$. Obviously, $S_e(a/b) = 1/S_e(b/a)$. This relationship is satisfied to about 1\% over the range considered. For the two largest $a/b$, the convergence is less, but likely better than 2\%. The values for $H_{\textrm{max}} = 0.02$ are plotted in Fig.~\ref{fig:rectangle-S}.  In these calculations, the far-field dipole temperature distribution was based on a strength $2S_e$ at a separation $b$. 

\begin{table}[tb]
\caption{Exterior shape factor $S_e$ for a rectangle of width $a$ and height $b$. For the interior, $S_i = a/b$. Far field has a dipole temperature distribution at radius~20.}\label{tab:rect-conv}
\centering
\setlength\tabcolsep{3pt}
\begin{tabular}{@{\hspace*{2pt}}lccccccc@{\hspace*{2pt}}}
\toprule
& \multicolumn{7}{c@{\hspace*{2pt}}}{$a/b$}\\
 \cmidrule(lr{2pt}){2-8}
$H_{\textrm{max}}$ & 0.25 & 0.5 & 1 & 2 & 4 & 8 & 16 \\
\midrule
0.05 & 0.7496 & 0.8584 & 0.9885 & 1.1377 & 1.3012 & 1.4706 & 1.6368 \\ 
0.02 & 0.7535 & 0.8637 & 0.9948 & 1.1462 & 1.3136 & 1.4908 & 1.6702 \\ 
\bottomrule
\end{tabular}%
\end{table}



\subsection{Regular \(K\)-gons}\label{sec:FEM-kgon}
The exterior shape factors were computed using the same procedure as for the rectangles (with accommodations for the $S^{2\mh2}$ hexagon, Fig.~\ref{fig:hex-flux-plots}, in which an isothermal edge adjoins the bottom edge). The exterior results are within 1.4\% of the interior FEM results.  The specific choice of outer boundary condition (isothermal, adiabatic, or dipole) affects the results. For the maximum radii and minimum mesh sizes achievable, isothermal outer boundaries run a fraction of a percent higher than dipole boundaries, and adiabatic boundaries run lower by 1--1.5\%.

The interior FEM calculations were straightforward, and comparison to known exact values and reciprocal relationships, together with grid convergence studies, places their accuracy at about 0.05\%. The reciprocal relationship~\cite{hersch1982harmonic} states that, when isothermal and adiabatic boundaries are interchanged, the resulting shape factor is the reciprocal of the original shape factor. Thus, in  Fig.~\ref{fig:hex-flux-plots} the following relationships are met, as is evident upon comparison of the respective flux plots: $S_p^{1\mh1} = 1/S_p^{2\mh1}$, $S_h^{1\mh1}  = 1/S_h^{2\mh2} $, $S_h^{1\mh1R} = 1/S_h^{3\mh1}$. The reciprocal values agree to 0.05\% to 0.1\%.

For \rule{0pt}{8pt}$K>4$, the interior flux has a weak singularity at corners that join isothermal and adiabatic segments, but grid refinement was sufficient to account for these.  For the hexagon with half of a side isothermal (Table~\ref{tab:FEM-exact}, last item), a stronger interior singularity is present at the transition point from isothermal to adiabatic (see \cite{lienhard2023potential}, \S6.2), and a numerical correction was applied using methods similar to Appendix~\ref{sec:rectangles}.


\section{Schwarz-Christoffel Mappings for Regular \(K\)-gons}\label{sec:SC-maps}
The interior Schwarz-Christoffel mapping, in general, is
\begin{equation}
 f(z) = c_1 \int_0^{z} \frac{dz}{(b_1-z)^{\mu_1}\cdots(b_K -z)^{\mu_K}} + c_2
\end{equation}
where $c_1$ and $c_2$ are constants~\cite{nehari}. The mapping of the interior of the unit disc to either the interior or the exterior of a regular $K$-gon uses as pre-vertices the $K$ roots of unity, $b_k = \exp(2\pi \mathrm{i} k/K)$. The exterior angles of the $K$-gon are $\pi\mu_i =2\pi/K$.  With
\begin{equation} 
\prod_{k=1}^K (b_k-z) = \prod_{k=1}^K \Bigl(\eu^{2\pi \mathrm{i} k/K}-z\Bigr) = \big(1-\cramped{z^K}\big) 
\end{equation}
 the mapping of the unit disc to the interior of a regular $K$-gon reduces to
\begin{equation}\label{eqn:disk-ngon-int}
f(z) =  \int_0^{z} \frac{dz}{\big(1 -\cramped{z^K}\big)^{2/K}}  
\end{equation}
where we take $c_1 = 1$ and $c_2 = 0$.

The mapping of the disc interior to the exterior of a polygon is
\begin{equation} 
 f(z) = c_3 \int_{z_0}^z \frac{(b_1-z)^{\mu_1}\cdots(b_K -z)^{\mu_K}\, dz}{z^2}, \quad z_0\neq 0 
\end{equation}
with $c_3$ a constant~\cite{nehari}. We have the same pre-vertices and exterior angles as for the interior case. With $c_3 =1$, the mapping of the disc to the exterior of the $K$-gon is 
\begin{equation}\label{eqn:disk-ngon-ext}
f(z) = \int_{z_0}^z \frac{\big(1-\cramped{z^K}\big)^{2/K}\, dz}{z^2}, \quad z_0\neq 0 
\end{equation}

Since the same points, the pre-vertices, are conformally mapped to the vertices of either the interior or exterior $K$-gon, and since isotherms map to isotherms and adiabats to adiabats, the boundary conditions are identical for the interior and exterior problems. It follows immediately see that the shape factor, a conformal invariant, is the same for the interior and the exterior problems on the $K$-gon.   (Note that we have not shown that the interior and exterior $K$-gons have the same dimensions or the same vertices---neither question is relevant to our result.)


\section{A Short Catalog of Results}\label{sec:catalog}
In this appendix, we provide a brief tabulation of shape factors for a few additional objects for which interior and exterior shape factors are equal.  This catalog is by no means exhaustive.

\begin{figure}[!t]
\begin{subfigure}{0.49\columnwidth}
\centering
\includegraphics[keepaspectratio=true,width=0.93\columnwidth]{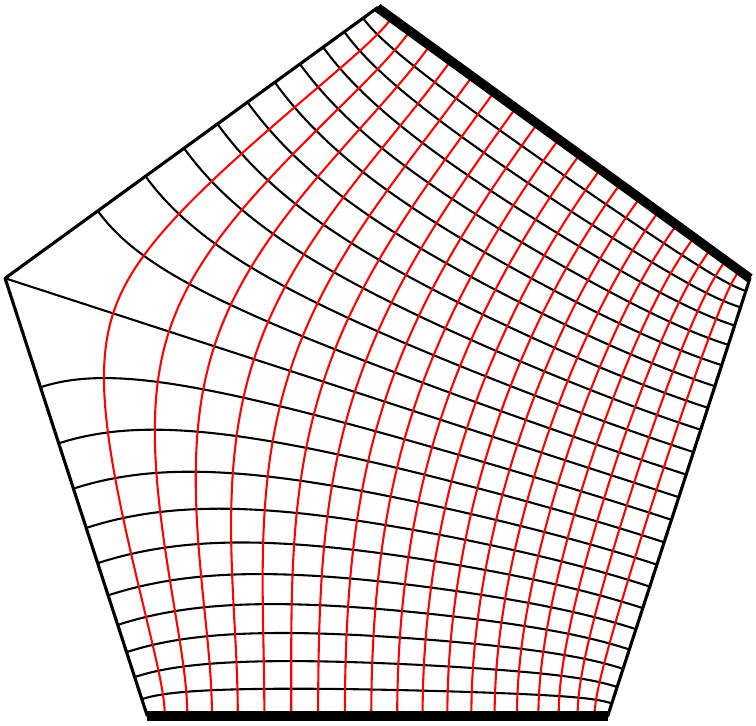}
\subcaption{$S_p^{1\mh1}=0.8963$\label{fig:pent-S11}}
\end{subfigure}
\begin{subfigure}{0.49\columnwidth}
\centering
\includegraphics[keepaspectratio=true,width=0.93\columnwidth]{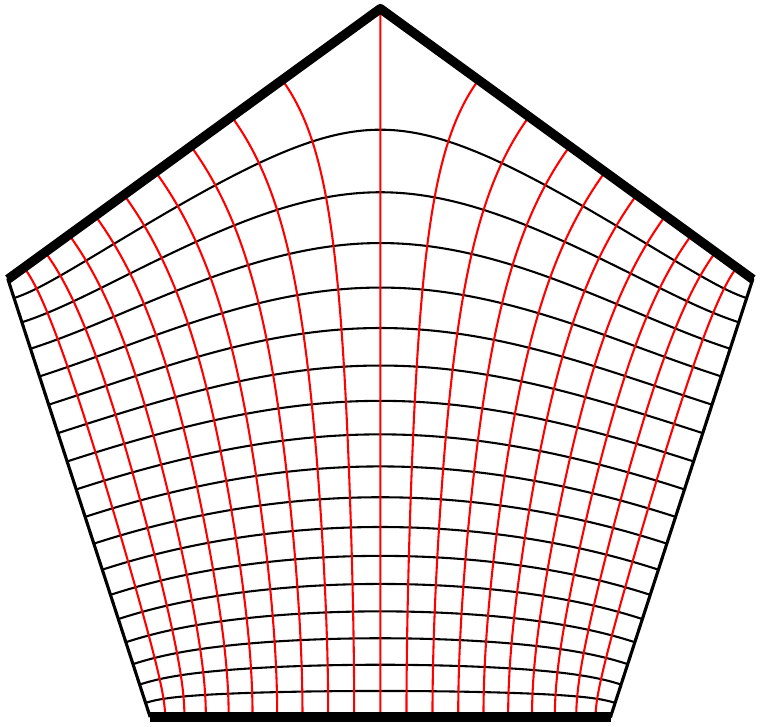}
\subcaption{$S_p^{2\mh1}= 1.1157$\label{fig:pent-S21}}
\end{subfigure}
\vskip10pt
\begin{subfigure}{0.49\columnwidth}
\centering
\includegraphics[keepaspectratio=true,width=0.93\columnwidth]{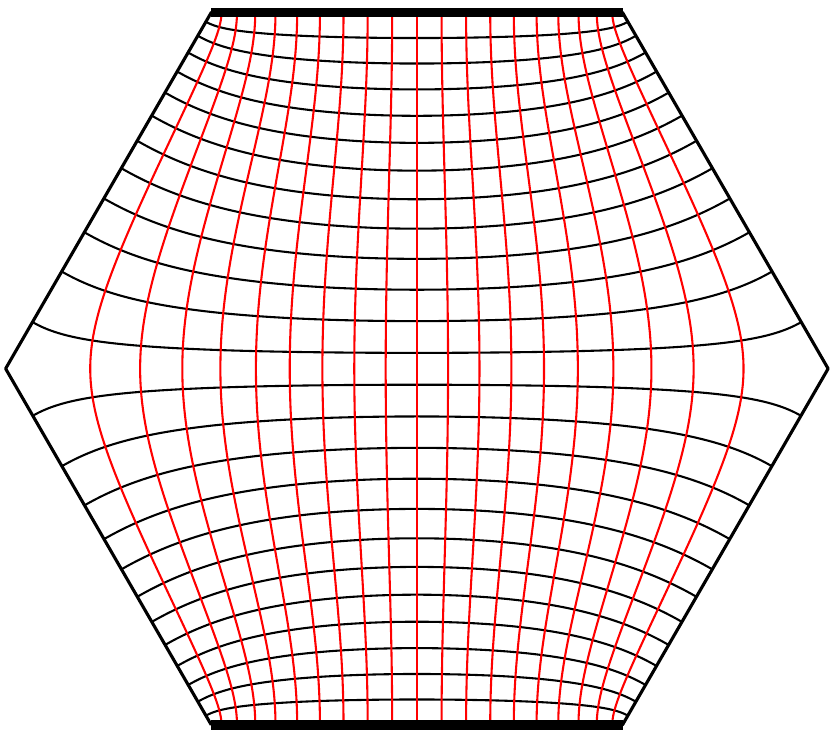}
\subcaption{$S_h^{1\mh1}=0.7815$\label{fig:hex-S11}}
\end{subfigure}
\begin{subfigure}{0.49\columnwidth}
\centering
\includegraphics[keepaspectratio=true,width=0.93\columnwidth]{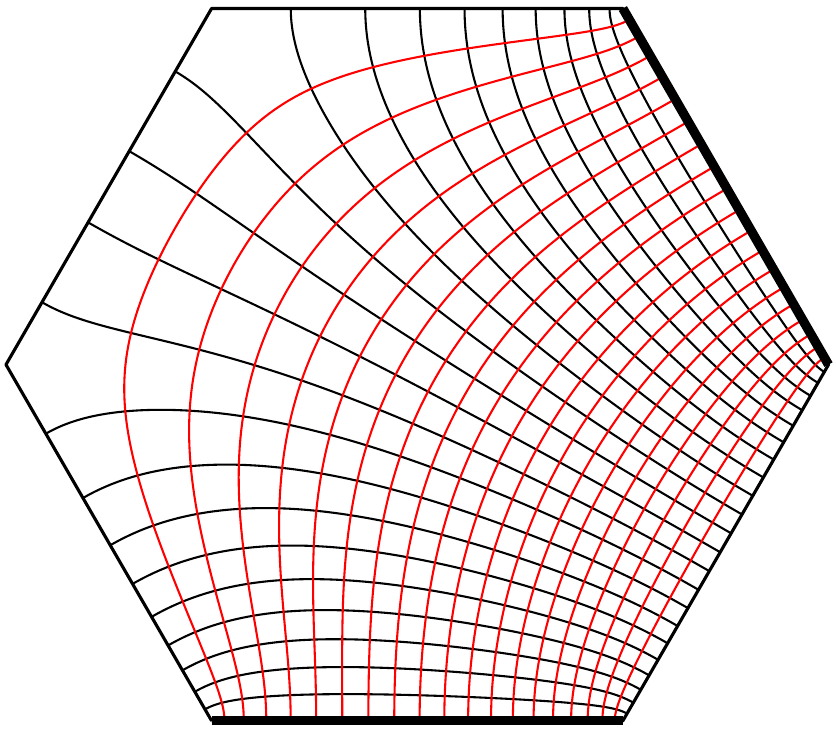}
\subcaption{$S_h^{1\mh1R}=0.8544$\label{fig:hex-S11R}}
\end{subfigure}
\vskip10pt
\begin{subfigure}{0.49\columnwidth}
\centering
\includegraphics[keepaspectratio=true,width=0.93\columnwidth]{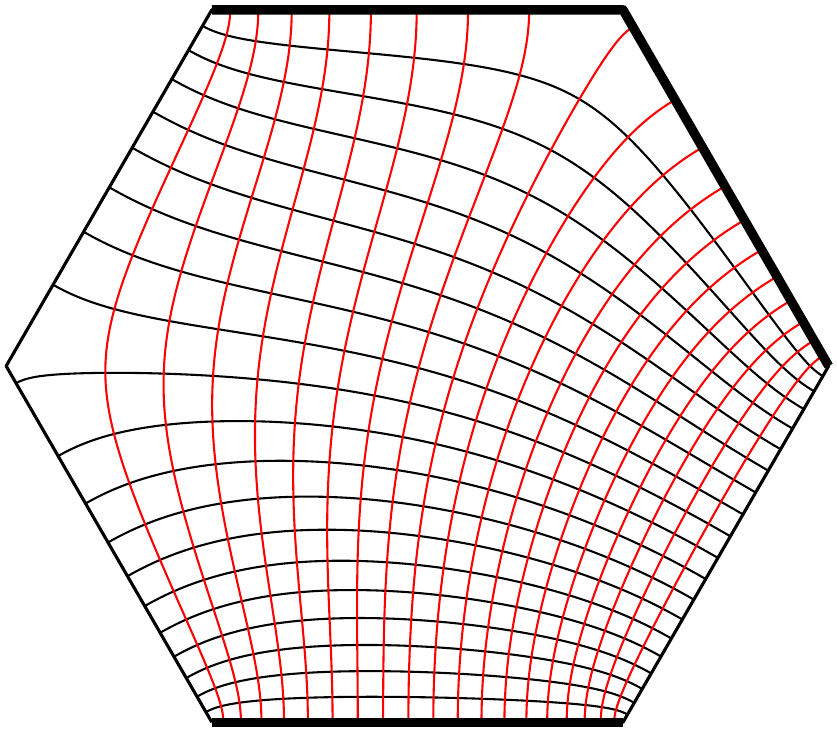}
\subcaption{$S_h^{2\mh1}=1$\label{fig:hex-S21}}
\end{subfigure}
\begin{subfigure}{0.49\columnwidth}
\centering
\includegraphics[keepaspectratio=true,width=0.93\columnwidth]{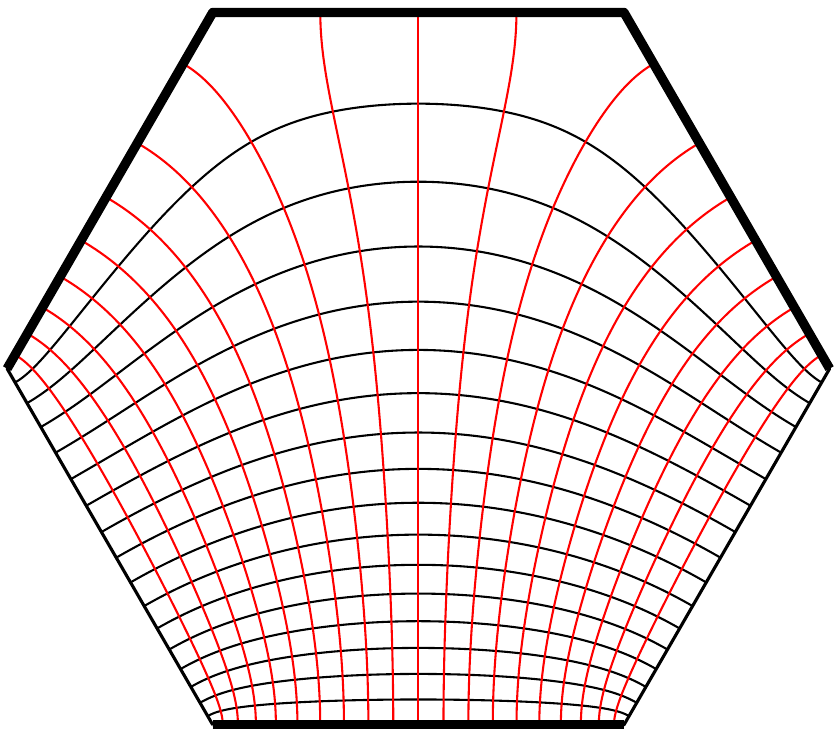}
\subcaption{$S_h^{3\mh1}=1.1699$\label{fig:hex-S31}}
\end{subfigure}
\vskip10pt
\begin{subfigure}{0.49\columnwidth}
\centering
\includegraphics[keepaspectratio=true,width=0.93\columnwidth]{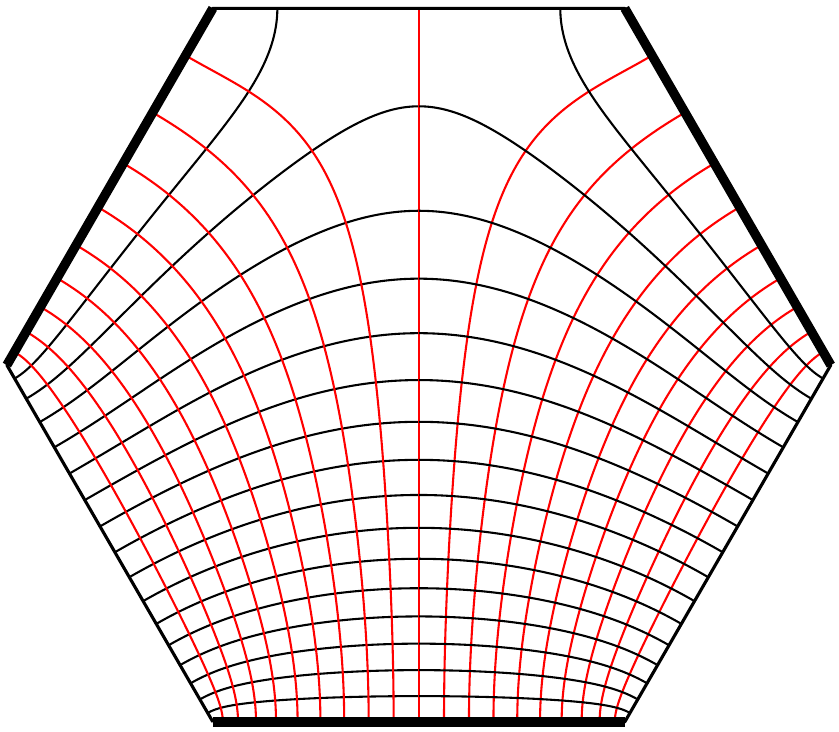}
\subcaption{$S_h^{2n\mh1}=2\big/\sqrt{3}$\label{fig:hex-S2n1}}
\end{subfigure}
\begin{subfigure}{0.49\columnwidth}
\centering
\includegraphics[keepaspectratio=true,width=0.93\columnwidth]{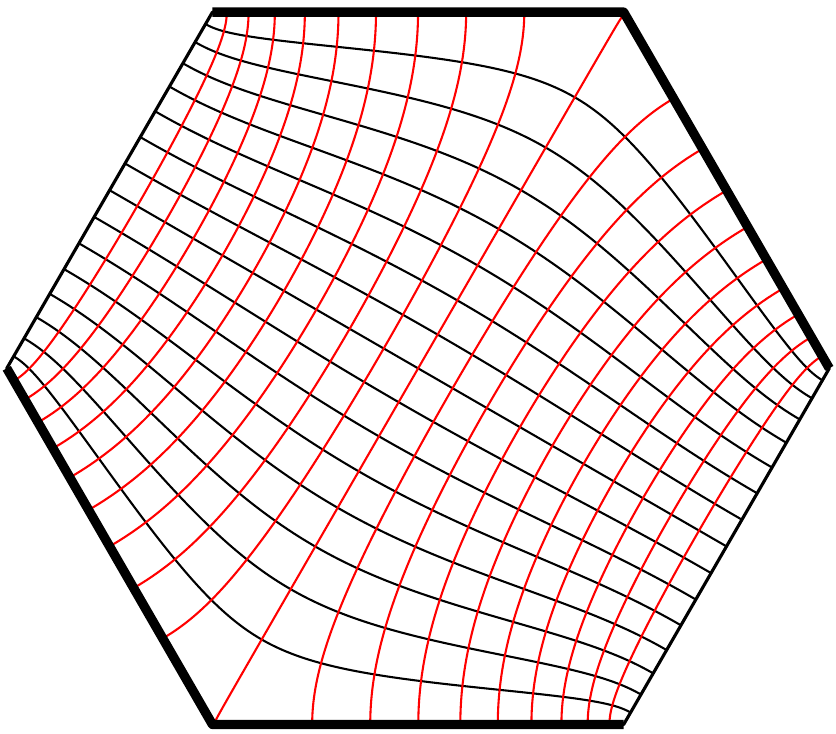}
\subcaption{$S_h^{2\mh2}=1.2790$\label{fig:hex-S22}}
\end{subfigure}
\caption{\label{fig:hex-flux-plots} Interior flux plots for all unique regular pentagons and hexagons. Black curves are isotherms, and red curves are adiabats. Isothermal sides are shown with thick lines. }
\end{figure}

\paragraph{Regular \(K\)-gons} Interior flux plots for all unique regular pentagon and regular hexagon problems are shown in Fig.~\ref{fig:hex-flux-plots}. The isothermal sides are indicated with heavy lines, and the other sides are adiabatic. In all cases, the bottom side has a different temperature than the top sides. The specific boundary conditions are apparent from the flux plots. Shape factors are given in the caption.  

The FEM results are accurate to 0.05\% or better. The values for $S_h^{2\mh1}$ and $S_h^{2n\mh1}$ are exact: the first case is a Yin-Yang body, and the second is from Hersch~\cite{hersch1982harmonic}). The exterior shape factors were computed, and in every case the value is within 1.4\% of the interior value; the difference is within the accuracy of the exterior computation. (FEM convergence is discussed in Appendix~\ref{sec:FEM-conv}.)  These results support the equality of interior and exterior shape factors for regular $K$-gons.  

\paragraph{Yin-yang bodies} A Yin-Yang body has an axis of symmetry about which isothermal and adiabatic boundaries are interchanged (\S\ref{sec:yiny}). Every Yin-Yang body is conformally equivalent to a square and has $S_i = S_e = 1$.

\begin{figure}[tb]
    \centering
    \includegraphics*[width=0.55\columnwidth]{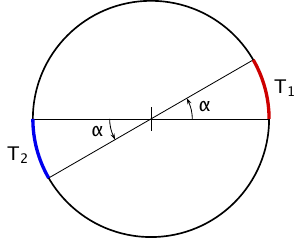}
    \caption{Disc with isothermal sectors of angle \(\alpha\).}
    \label{fig:disc-alpha}
\end{figure}

\paragraph{Discs} A disc with symmetric isothermal angular sectors is shown in Fig.~\ref{fig:disc-alpha}.  The shape factors for several angles are given in Table~\ref{tab:disc-alpha}; and the angles that yield several shape factors are given in Table~\ref{tab:disc-S}. For all discs, $S_i \equiv S_e$.

The case $\alpha = 30$\textdegree\ has the analytical solution $S = 1/\sqrt{3}$~\cite{hersch1982harmonic}; and, because the case $\alpha = 150$\textdegree\ simply interchanges the boundary conditions, reciprocity~\cite{hersch1982harmonic} shows it to have the analytical solution $S = \sqrt{3}$.  Similarly, the shape factor for $\alpha$ is the reciprocal of the shape factor for $180\textrm{\textdegree} - \alpha$.  The case $\alpha = 90$\textdegree\ is a Yin-Yang body with $S =1$.

These results were computed using Driscoll's \textit{Schwarz-Christoffel Toolbox}~\cite{driscollsctb}.  A conformal mapping from the unit disc to a rectangle of varying aspect ratio was sought because the shape factor of the rectangle is equal to the aspect ratio and the same value must apply to the disc. The pre-vertices on the unit disc were calculated, while forcing the conformal center of the map to the center of the rectangle.  The results are accurate to the number of digits shown in Table~\ref{tab:disc-alpha}.

\begin{table}[tbh]
\caption{Shape factor for various angles on a symmetric sector disc. (Numerical details in Appendix~\ref{sec:FEM-conv}.)}\label{tab:disc-alpha}
   \centering
    \setlength\columnsep{2pt}
    \begin{tabular}{lcccccc}
    \toprule
    $\bm{\alpha}$\rlap{\textdegree} & 90.00 &  75.00  & 60.00  & 45.00 & 30.00  & 15.00 \\
    $\bm{S}$ & 1.0000 & 0.8865 &  0.7817 &  0.6806 & $1/\sqrt{3}$ & 0.4595    \\
    \midrule
    $\bm{\alpha}$\rlap{\textdegree}  & 90.00 & 105.00 & 120.00  & 135.00 & 150.00 & 165.00\\
    $\bm{S}$    & 1.0000  & 1.1280 & 1.2793 & 1.4692 &$\sqrt{3}$ & 2.1761\\
    \bottomrule
    \end{tabular}
\end{table}

\begin{table}[!htb]
\caption{Angle producing various shape factors on a symmetric sector disc.}\label{tab:disc-S}
   \centering
    \setlength\columnsep{2pt}
    \begin{tabular}{lccccc}
    \toprule
    $\bm{S}$ & 1.0000 & 1.5000 &  2.0000 &  3.0000 & 5.0000    \\
    $\bm{\alpha}$\rlap{\textdegree} & 90.00 &  137.07 & 160.24  & 175.88& 179.82 \\
    \midrule
    $\bm{S}$    & 1.0000  & 0.7500 & 0.5000 & 0.3000 & 0.2000 \\
    $\bm{\alpha}$\rlap{\textdegree}  & 90.00 & 55.32 & 19.76  & 2.439 & 0.01779 \\
    \bottomrule
    \end{tabular}
\end{table}


\bibliographystyle{asmejour}
\bibliography{int_ext}

\end{document}